\documentclass[nojss]{jss}
\usepackage{amsmath,amssymb,amsfonts,thumbpdf}
\usepackage{natbib}
\usepackage{multirow}
\usepackage{algorithm}
\usepackage{algpseudocode}
\usepackage{placeins}
\usepackage{orcidlink}
\usepackage{caption}
\usepackage{xcolor}

\author{
  Mattias Wetscher~\orcidlink{0000-0002-9982-3001}\\University of Innsbruck \And
  Johannes Seiler~\orcidlink{0000-0001-5714-9234}\\University of Innsbruck \AND
  Reto Stauffer~\orcidlink{0000-0002-3798-5507}\\University of Innsbruck \And
  Nikolaus Umlauf~\orcidlink{0000-0003-2160-9803}\\University of Innsbruck
}
\Plainauthor{Mattias Wetscher, Johannes Seiler, Reto Stauffer, Nikolaus Umlauf}

\title{Stagewise Boosting Distributional Regression}

\Shorttitle{Stagewise Boosting Distributional Regression}

\Abstract{
Forward stagewise regression is a simple algorithm that can be used to estimate regularized 
models. The updating rule adds a small constant to a regression 
coefficient in each iteration, such that the underlying optimization problem is solved slowly with 
small improvements. This is similar to gradient boosting, with the essential difference 
that the step size is determined by the product of the gradient and a step length parameter in the 
latter algorithm. One often overlooked challenge in gradient boosting for distributional regression
is the issue of a vanishing small gradient, 
which practically halts the algorithm's progress. We show that gradient boosting in this case
oftentimes results in suboptimal models, especially for 
complex problems certain distributional parameters are never updated due to the vanishing gradient. 
Therefore, we propose a stagewise boosting-type algorithm for distributional regression, combining stagewise regression ideas with gradient boosting.
Additionally, we extend it with a novel regularization method, correlation filtering, 
to provide additional stability when the problem involves a large number of covariates. Furthermore, the algorithm includes best-subset selection for parameters and can be applied to big data problems by leveraging stochastic approximations of the updating steps. Besides the advantage of processing large datasets, the stochastic nature of the approximations can lead to better results, especially for complex distributions, by reducing the risk of being trapped in a local optimum. The performance of our proposed stagewise boosting distributional regression approach is investigated in an extensive simulation study and by estimating a full probabilistic model for lightning counts with data of more than 9.1 million observations and 672 covariates.
}

\Plainkeywords{Distributional Regression, Stagewise Regression, Gradient Boosting,
  Variable Selection, Correlation Filtering, Batchwise Updating}

\Address{
  Mattias Wetscher, Johannes Seiler, Reto Stauffer, Nikolaus Umlauf\\
  Department of Statistics\\
  Faculty of Economics and Statistics\\
  Universit\"at Innsbruck\\
  6020 Innsbruck, Austria\\
  E-mail: \email{Mattias.Wetscher@uibk.ac.at}\\
  \phantom{E-mail: }\email{Johannes.Seiler@uibk.ac.at},\\
  \phantom{E-mail: }\email{Reto.Stauffer@uibk.ac.at},\\
  \phantom{E-mail: }\email{Nikolaus.Umlauf@uibk.ac.at}\\
  URL: \url{https://nikum.org/},\\
  \phantom{URL: }\url{https://retostauffer.org/}\\
}

\begin{document}

\def\spacingset#1{\renewcommand{\baselinestretch}%
{#1}\small\normalsize} \spacingset{1}

\section{Introduction}

Modern regression models can  not only provide estimates for the expectation, but full probabilistic 
predictions, which are particularly important in numerous applications, e.g., for the prediction of severe weather using a complex count model for the number of lightning strikes as demonstrated in this article
\citep[similar models where used in,][]{sdr:Simon+Fabsic+Mayr+Umlauf+Zeileis:2018, sdr:Simon+Mayr+Umlauf+Zeileis:2019}.
A well-known model class for such probabilistic settings is the generalized additive model
for location,  scale and shape
\citep[GAMLSS;][]{sdr:Rigby+Stasinopoulos:2005,sdr:Klein+Kneib+Lang:2015},
where each parameter of the response distribution is modeled by covariates.
In order to estimate a well-calibrated forecast, a stable estimation algorithm is needed for
GAMLSS, which is also able to perform variable selection at the same time. A common choice in
such situations is the gradient boosting algorithm for GAMLSS \citep{sdr:Hofner2021}, which
can perform variable selection for parameters of the response distribution in high dimensions, i.e.,
even if the problem contains more covariates than observations.

Gradient boosting in the aforementioned framework iteratively improves the regression coefficients, thereby fitting the covariates to the negative gradient of the log-likelihood function
with respect to the linear predictors and only updating the best performing covariate, either for
each parameter of the distribution, also called \emph{cyclical update}, or only the best performing
parameter, called \emph{non-cyclical update} \citep{sdr:Thomas2018GradientBF}. Essentially,
this updating scheme is a coordinate descent step, where in each iteration only the covariate with the largest (in absolute value) partial derivative gets updated. It is performed at a
learning rate such that the objective function improves very slowly from iteration to iteration,
allowing for stable estimates with shrinkage.

\begin{figure}[ht!]
\centering
\includegraphics[width=0.8\textwidth]{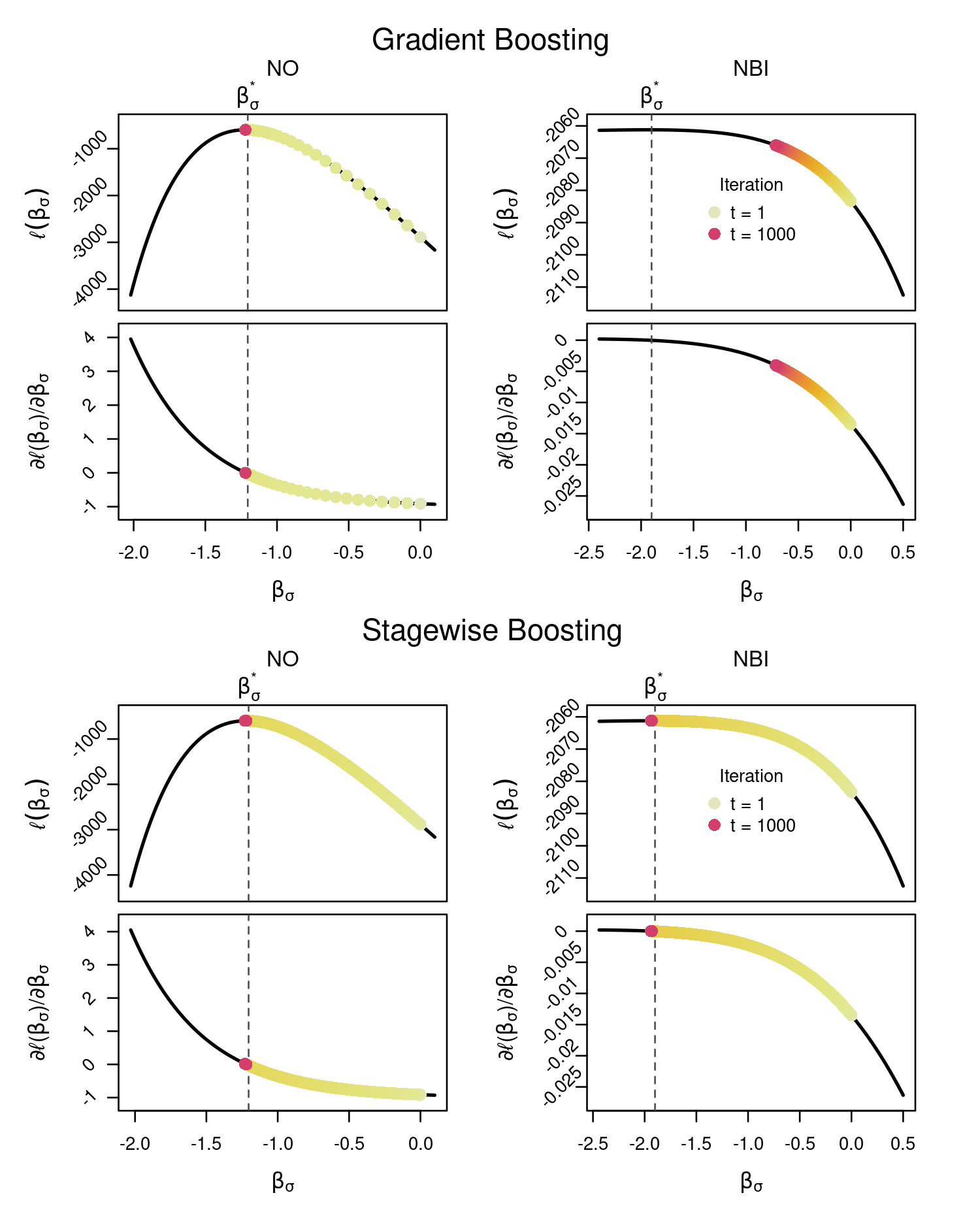}
\caption{\label{fig:gradproblem} Illustration of the vanishing gradient problem in boosting 
  when estimating distributional regression models. The first and second row show an updating procedure
  based on the gradient. Row three and four show the stagewise counterpart where a constant updating
  step size $\epsilon = 0.01$ is used. The first and third row show the (marginal) log-likelihood functions
  $\ell( \beta_{\sigma} )$ of the $\texttt{NO}(\mu = \beta_{\mu}, \log(\sigma) = \beta_{\sigma})$ in the
  left column and the $\texttt{NBI}(\log(\mu) = \beta_{\mu}, \log(\sigma) = \beta_{\sigma})$
  in the right column. The second and fourth row show the corresponding gradient information.
  The true parameters ($\beta_{\sigma}^\star$) are indicated by a vertical gray dashed line.}
\end{figure}
A particular disadvantage of gradient boosting is the search for the optimal stopping iteration of 
the algorithm, since too many iterations lead to overfitting. This problem is usually solved by 
performing cross-validation or a computational more efficient method, such as probing \citep{Thomas2017_2}, which expands the original data with independent ``shadow variables''. If such a shadow variable is selected, the updating algorithm stops.
However, in the context of distributional regression, this
procedures can easily lead to suboptimal models, especially when gradient boosting is applied to 
complex distributions, since the updating scheme depends heavily on the gradient information.
This is the case when the iterations for some parameters quickly fall into 
regions of the objective function that are particularly flat, i.e., when the gradients become 
vanishingly small. As a result, some parameters of the distribution may never be updated for both procedures,
and using cross-validation, the number of boosting iterations would be extremely high to find the 
optimum.

The problem of parameters that are never selected for updating is discussed in
\citet{sdr:2102.09248}, where an adaptive step length is proposed and analytical optimal step length 
parameters are provided for the normal location scale model. The selection of the optimal 
step length is based on minimizing the negative log-likelihood for each base learner. This step length is then multiplied by a 
penalty $\lambda$ (e.g., $\lambda = 0.1$) to again force slow improvements in the objective 
function. Despite these efforts, the utilization of gradients in this technique may still result in 
the occurrence of vanishing gradients in complex distributional regression settings, leading to a 
selection problem with a possibly high proportion of false negatives.

To address the issues related to vanishing gradients in boosting distributional regression models, we
propose to adapt forward stagewise regression \citep{Tibshirani2015}. This updating 
scheme is very similar to gradient boosting, but only uses information of the sign of the gradient
to update coefficients at a small constant rate $\epsilon$ (e.g., $\epsilon = 0.01$).
\citet{Tibshirani2015} shows the efficiency of the approach in a general framework for convex optimization.
Since the step length of the update is kept constant, a stagewise algorithm tailored for
distributional regression does not suffer from the vanishing gradient, and a more balanced choice of
covariates for the update is possible if $\epsilon$ is sufficiently small.

\pagebreak

The vanishing gradient problem in distribution regression is illustrated in 
Figure~\ref{fig:gradproblem}, where we compare the convergence behavior of the parameters of a 
gradient-based updating scheme and our proposed stagewise boosting updating scheme for the normal 
distribution (\code{NO}($\mu$, $\sigma$)) and the negative binomial distribution of type one 
(\code{NBI}($\mu$, $\sigma$)). Let us assume that the algorithms for both the \code{NO} and 
the \code{NBI} distribution have already approximately reached the optimum for the first 
parameter $\mu$, i.e., that no further update is required for $\mu$. In this case, the algorithms
will only update the $\sigma$ parameter. This behavior becomes
apparent in simple numerical examples with the \code{NBI} distribution. Nonetheless, it is 
noteworthy that employing the \code{NO} distribution can also result in similar convergence 
patterns. As can be seen in Figure~\ref{fig:gradproblem}, the subsequent update iterations for the second
parameter $\sigma$ of the \code{NO} distribution converge very quickly to the optimum
$\beta^\star_{\sigma}$ with both methods, the gradient information only becomes very
small near the optimum. However, the subsequent gradient boosting updates for parameter $\sigma$ of
the \code{NBI} distribution do not reach the optimum after 1000 iterations and are 
already slowed down after the first few iterations due to the very small gradient,
as indicated by the updating points.
In contrast, our proposed stagewise updating for distributional regression does not
contain the full information of the gradient and instead uses a (semi-)constant step length so
that it can converge to the true parameter $\beta^\star_{\sigma}$ after a few iterations.
This behavior of gradient boosting algorithms for distributional regression implies the following: In a larger regression setting with 
a number of covariates, some coefficients may never be updated (i.e., remain zero) because distribution parameters with larger 
gradients dominate parameters with vanishingly small gradients in the updating process. In this case, users would have to run the 
boosting algorithm for a very long time to detect relevant variables. In practice, however, this problem goes unnoticed because users 
typically only look at the evolution of the log-likelihood to see if the boosting algorithm has ``converged''. And since the gradients 
and update steps are very small, one concludes that no further improvement in the log-likelihood can be achieved. This behavior is well 
illustrated in Figure~\ref{fig:gradproblem} by the colored update steps, which show a sharp break (far away from the optimum
$\beta_{\sigma}^\star$), so to speak, resulting in a very flat (almost horizontal) line when looking at the log-likelihood curve over 
iterations. Similarly, this behavior is usually not seen when looking at individual coefficient paths. Finally, this effect naturally 
leads to high false negative rates (weak parameters) and high false positive rates (dominant parameters) in settings with a large 
number of potential variables to choose from (see simulation Section~\ref{sec:simulation}).

In order to address the aforementioned challenges associated with the use of gradient boosting for 
distributional regression, this paper presents a novel approach to distributional regression by 
adapting and expanding upon the forward stagewise algorithm by
\begin{itemize}
  \item best-subset selection of distributional parameters in each iteration with a semi-constant step length,
  \item simple selection of stopping iterations by a novel correlation filtering approach,
  \item a batchwise variant of the algorithm for big data, based on stochastic approximations of the updating steps,
    mitigating the likelihood of being trapped in local optima.
\end{itemize}
Furthermore, we present an extensive simulation study in which we test the proposed methods in comparison to established state-of-the-art methods in highly challenging scenarios.
We use two two-parameter distributions,
a normal ($\texttt{NO}(\mu, \sigma)$) and a gamma distribution ($\texttt{GA}(\mu, \sigma)$),
and a three-parameter distribution, the
zero-adjusted negative binomial distribution of type I ($\texttt{ZANBI}(\mu, \sigma, \nu)$), with different 
numbers of uninformative variables (30 and 100) for each distributional parameter and different 
degrees of  multicollinearity between the variables.

To emphasize the usefulness of the approach, our scalable version of the algorithm is
illustrated by a prediction problem of lightning counts using a high-dimensional dataset with
approximately $9.1$ million observations and $672$ variables. The prediction is based on the $
\texttt{ZANBI}$ distribution, where each of the three distribution parameters is modeled by the 
variables selected by our algorithm.

The remainder of this paper is organized as follows: In Section~\ref{sec:modelspec} we give
a brief overview of distributional regression models. Gradient boosting for distributional regression
is described in Section~\ref{sec:gradboost}. Stagewise regression in general as well as our novel variants for distributional regression using semi-constant step length, best-subset updating, correlation filtering and batchwise estimation are presented in Section~\ref{sec:stag}.
In Section~\ref{sec:simulation} we present the results of the simulation study, in
Section~\ref{sec:flash} we demonstrate the application of our methods to the complex
problem of modeling lightning counts.

\section{Distributional Regression} \label{sec:modelspec}

The underlying idea of distributional regression
\citep{sdr:Rigby+Stasinopoulos:2005, sdr:Klein+Kneib+Klasen+Lang:2015, umlauf2018b}
is to model all parameters of an arbitrary parametric response distribution
(rather than just the mean) through covariates. In the following, we briefly introduce the
framework and describe classic maximum likelihood estimation of the parameters.

\subsection{Model Specification}\label{modspec}

For a dataset with $i = 1, \ldots, n$ observations, where $y_i$ denotes the response (which may
also be non-continuous or multivariate) and $\mathbf{x}_i$ available covariate information,
we assume conditional independence of the individual response observations given covariates.
Specifically, let
\begin{equation*} \label{eqn:dreg}
y_i \sim \mathcal{D}\left(\theta_{1}(\mathbf{x}_i), \dots,  \theta_{K}(\mathbf{x}_i)\right),
\end{equation*}
where $\mathcal{D}(\cdot)$ represents a distribution with $K$ parameters
$\theta_{ik} \equiv \theta_{k}(\mathbf{x}_i)$, $k = 1, \ldots, K$, that are linked to additive
predictors $\eta_{ik} \equiv \eta_{k}(\mathbf{x}_i)$ using known monotonic and twice
differentiable functions $h_{k}(\cdot)$, 
\begin{equation*} 
h_{k}(\theta_{ik}) = \eta_{ik} = \mathbf{x}_{ik}^\top\boldsymbol{\beta}_k.
\end{equation*}
Here, $\mathbf{x}_{ik}$ is a row of the $n \times J_k$ predictor specific design matrix
$\mathbf{X}_k$ and the parameters
$\boldsymbol{\beta}_k = (\boldsymbol{\beta}_{0k}, \ldots, \boldsymbol{\beta}_{J_kk})^\top$ are
regression coefficients. In matrix notation, the predictors are written as
\begin{equation*}
\boldsymbol{\eta}_k=(\eta_{1k},\ldots,\eta_{nk})^\top = \mathbf{X}_k\boldsymbol{\beta}_k,
\end{equation*}
which is the fully parametric case of the GAMLSS \citep{sdr:Rigby+Stasinopoulos:2005}.


\subsection{Classic Likelihood Estimation}

Parameter estimation is based on maximizing the log-likelihood function
\begin{equation} \label{eqn:loglik}
\ell(\boldsymbol{\beta} ; \mathbf{y}, \mathbf{X}) =
  \sum_{i = 1}^n \log \, d_y\left(y_i; \theta_{i1} = h_1^{-1}(\eta_{i1}), \ldots,
  \theta_{iK} = h_K^{-1}(\eta_{iK})\right),
\end{equation}
where $d_y( \cdot )$ is the density function and $\mathbf{y} = (y_1, \ldots, y_n)^\top$ the response vector,
$\boldsymbol{\beta} = (\boldsymbol{\beta}_1^\top, \ldots, \boldsymbol{\beta}_K^\top)^\top$ the
stacked vector of regression coefficients to be
estimated and $\mathbf{X} = (\mathbf{X}_1, \ldots, \mathbf{X}_K)$ represents the full covariate
data matrix.

For the maximization of \eqref{eqn:loglik}, \citet{sdr:Rigby+Stasinopoulos:2005} use a modified
backfitting algorithm based on iteratively reweighted least squares \citep[IWLS;][]{sdr:Gamerman:1997}.
Similarly, \citet{sdr:Umlauf+Klein+Zeileis:2018} use IWLS based iterations for both,
maximum likelihood and full Bayesian estimation by Markov chain Monte Carlo (MCMC) simulation.

An important aspect in estimating distributional regression models often involves variable selection. Classical maximum likelihood methods, due to their cyclic updating nature, are not suitable for selection and regularization. In contrast, boosting updates inherently perform variable selection through early stopping, rendering it more suitable for this purpose.
Another shrinkage method is presented by
\citet{sdr:Groll+Hambuckers+Kneib+Umlauf:2019} based on $L_1$-type regularization in the context
of distributional regression for metric covariates and both group and fused-LASSO for categorical
variables. The major drawback here, however, is that the algorithm allows little flexibility and is
additionally numerically very expensive as the shrinkage parameters must be determined
using a grid search. This makes the method unattractive for many applications, e.g., models for
very large data like in our application (see Section~\ref{sec:flash}).
Therefore, boosting-type algorithms are currently considered the leading methods for
variable selection in the context of distributional regression. This is largely attributed
to their exceptional numerical stability, which makes them particularly effective for this purpose,
as described in more detail in the following section.

\section{Gradient Boosting} \label{sec:gradboost}

Gradient boosting is a widely used supervised machine learning technique that is particularly effective in constructing predictive models, including distributional regression models.
It is a type of boosting
algorithm that aims to improve the prediction of the model by iteratively minimizing the
residuals or a general loss function in very small steps.
In contrast to the traditional backfitting step, which updates every coefficient in each iteration,
gradient boosting for distributional regression follows a different approach by only updating the best-performing
regression coefficient overall (non-cyclical update) or the best-performing regression coefficient
for each distributional parameter (cyclical update) in each iteration
\citep{sdr:Hofner2021, sdr:Thomas2018GradientBF}\footnote{\citealp{sdr:Hofner2021} and \citealp{sdr:Thomas2018GradientBF} formulate gradient boosting with more general effects, i.e., baselearners and in the context of minimizing a loss function. Instead of the latter we present it in the context of maximizing a log-likelihood.}.
The nature of the iterative updating also makes gradient boosting a powerful technique for modeling complex
data even with only limited sample size.

\subsection{Cyclical Gradient Boosting} \label{sec:cyclic}

For the following assume that each covariate in $\mathbf{x}_i$ is standardized. To illustrate the concept of cyclical gradient boosting \citep{Hofner2014},
let us consider a linear model given by
$$
y_i = \mathbf{x}_i^\top\boldsymbol{\beta} + \varepsilon_i \quad \text{and} \quad
  \varepsilon_i \sim \mathcal{N}(0, \sigma^2),
$$
which can be seen as a special case of the more general distributional regression model:
$$
y_i \sim \mathcal{D}\left( y_i; \theta_{i1} = \eta_{i1}, \theta_{i2} = \exp(\eta_{i2}) \right),
$$
where $\mathcal{D}( \cdot )$ represents the two-parameter normal distribution
$\mathcal{N}(\mu, \sigma^2)$, and $\theta_{i1}$ and $\theta_{i2}$ correspond to the mean
parameter $\mu$ and variance parameter $\sigma^2$, respectively. In the homoskedastic case,
we have $\exp(\eta_{i2}) = \sigma^2 \,\, \forall \, i$. In the full distributional
regression setting, we model $\exp(\eta_{i2}) = \sigma_i^2$, hence, we specify the
linear predictors $\eta_{i1} = \mathbf{x}_{i1}^\top\boldsymbol{\beta}_1$ and
$\eta_{i2} = \mathbf{x}_{i2}^\top\boldsymbol{\beta}_2$.

In gradient boosting for distributional regression the intercepts, for the linear model
$\beta_{01}$ and $\beta_{02}$, are initialized by its corresponding maximum likelihood estimates,
which is usually relatively straightforward to compute. All other coefficients are initialized as zero.

Then, starting from the initial values
$$
\boldsymbol{\theta}_k^{[0]} = h_k^{-1}\left( \boldsymbol{\eta}_k^{[0]} \right) =
  h_k^{-1}\left( \mathbf{X}_k\boldsymbol{\beta}_k^{[0]} \right),
$$
in each iteration $t = 1, \ldots ,T$, of the boosting algorithm, the coefficients
$\boldsymbol{\beta}_1^{[t]}$ and $\boldsymbol{\beta}_2^{[t]}$ are improved sequentially and slowly. Starting from $\boldsymbol{\beta}_1^{[t]}$, fit each column of the predictor-specific model matrix $\mathbf{X}_1$, denoted by $({\mathbf{X}_{1}})_{\cdot j_1}$, to the gradient vector
\begin{equation*} 
\mathbf{g}_1 =
 \left(  \frac{\partial \, \log \, d_y(y_i; \theta_{i1}^{[t-1]}, \theta_{i2}^{[t-1]})}{\partial \eta_1}  \right)_{i=1,\dots,n}
\end{equation*}
and select the coefficient to be updated according to the residual sum of squares (RSS) criterion 
$$
j_1^* = \underset{{j_1=1,\dots,J_1}}{\mathrm{argmin}}
  \left\| \mathbf{g}_1-({\mathbf{X}_{1}})_{\cdot j_1} c_{j_1}   \right\|_2^2 ,
$$ 
where $c_{j_1}$ represents the coefficient of the least squares fit of column $({\mathbf{X}_{1}})_{\cdot j_1}$ with the current gradient vector $\mathbf{g}_1$.
The updating step then is
\begin{equation*} \label{eqn:upstep1}
\boldsymbol{\beta}_{1}^{[t]} = \boldsymbol{\beta}_{1}^{[t-1]} + \epsilon \cdot c_{j_{1}^*} \cdot \mathbf{e}_{j_1^*},
\end{equation*}
where $\epsilon$ is the step length control parameter (e.g., set to $\epsilon = 0.1$) and
$\mathbf{e}_{j_1^*}$ is an index vector consisting of zeros, except for the position $j_1$, where it contains a one.
This updating step moves the coefficients slowly in the direction of $\mathbf{g}_1$, which is
similar to a coordinate descent step and computationally efficient, since least squares
estimation is fast and numerically stable.
Secondly, the process is repeated with $\boldsymbol{\beta}_2^{[t]}$, but the updated version
$\boldsymbol{\beta}_1^{[t]}$ is used, i.e., each column of $({\mathbf{X}_{2}})_{\cdot j_2}$ is
fitted to the gradient vector
\begin{equation*} 
\mathbf{g}_2 =
 \left(  \frac{\partial \, \log \, d_y(y_i; \theta_{i1}^{[t]}, \theta_{i2}^{[t-1]})}{\partial \eta_2}  \right)_{i=1,\dots,n}.
\end{equation*}
Then, $j_2^*$  is determined with the respective RSS-criterion and the update

\begin{equation*} 
\boldsymbol{\beta}_{2}^{[t]} = \boldsymbol{\beta}_{2}^{[t-1]} + \epsilon \cdot c_{j_{2}^*} \cdot \mathbf{e}_{j_2^*},
\end{equation*}
is carried out. 

In more generality, with $K$ distibutional paramerters, the updating is sketched as follows:
\begin{align*}
\mathbf{g}_1 &=
 \left(\frac{\partial \,\log\, d_y(y_i,\mathbf{x}_i, (\boldsymbol{\beta}_1^{[t-1],\top}, \ldots, \boldsymbol{\beta}_K^{[t-1],\top})^\top   )}{\partial{\eta_1}}\right)_{i=1,\dots,n} &\underset{\text{update}}{\rightarrow}& \boldsymbol{\beta}_1^{[t]}\\
\mathbf{g}_2 &=
 \left(\frac{\partial \,\log\, d_y(y_i,\mathbf{x}_i, (\boldsymbol{\beta}_1^{[t],\top},\boldsymbol{\beta}_2^{[t-1],\top}, \ldots, \boldsymbol{\beta}_K^{[t-1],\top})^\top   )}{\partial{\eta_2}}\right)_{i=1,\dots,n} &\underset{\text{update}}{\rightarrow}& \boldsymbol{\beta}_2^{[t]}\\
& &\vdots& \\
\mathbf{g}_K &=
 \left(\frac{\partial \,\log\, d_y(y_i,\mathbf{x}_i, (\boldsymbol{\beta}_1^{[t],\top}, \ldots,\boldsymbol{\beta}_{K-1}^{[t],\top},\boldsymbol{\beta}_K^{[t-1],\top} )^\top   )}{\partial{\eta_K}}\right)_{i=1,\dots,n} &\underset{\text{update}}{\rightarrow}& \boldsymbol{\beta}_K^{[t]}.
\end{align*}

\subsection{Non-Cyclical Gradient Boosting} \label{sec:noncyclic}

Instead of cyclically updating all parameters, a variation of the gradient boosting algorithm focuses on updating only the best model term. This is done by selecting among the tentative updates
\begin{equation} 
\boldsymbol{\beta}_{k}^{[t]} = \boldsymbol{\beta}_{k}^{[t-1]} + \epsilon \cdot c_{j_{k}^*} \cdot \mathbf{e}_{j_k^*},
\end{equation}
with 
$$
j_k^* = \underset{{j_k=1,\dots,J_k}}{\mathrm{argmin}}
  \left\| \mathbf{g}_k-({\mathbf{X}_{k}})_{\cdot j_k} c_{j_k}   \right\|_2^2 ,
$$ 
only the one that maximizes the log-likelihood most
$$
k^* = \underset{{k=1,\dots,K}}{\mathrm{argmax}}~ \ell \left(\left(\boldsymbol{\beta}_1^{[t-1]\top},
  \ldots,\boldsymbol{\beta}_{k}^{[t]\top}, \dots \boldsymbol{\beta}_K^{[t-1]\top}\right)^\top ;
  \mathbf{y}, \mathbf{X}\right).
$$
The tentative updates are computed analogous to the cyclical updating but the respective column 
of the
predictor-specific model matrices $\mathbf{X}_k$ are fitted to a not updated version of the gradient,
\begin{equation} \label{eqn:gradvec}
\mathbf{g}_k =
 \left(  \frac{\partial \, \log \, d_y(y_i; \theta_{i1}^{[t-1]}, \ldots, \theta_{iK}^{[t-1]})}{\partial \eta_k}  \right)_{i=1,\dots,n}
\end{equation}
The final update in iteration $t$ is only carried out in the $k^*$-th distributional parameter
and for the cases $k \neq k^*$, the coefficients remain unchanged, i.e.,
$\boldsymbol{\beta}_{k}^{[t]} = \boldsymbol{\beta}_{k}^{[t-1]}$. In contrast to cyclic updating, this so-called non-cyclic updating method has proven advantages and is thoroughly investigated in \citet{sdr:Thomas2018GradientBF}.

\subsection{Remarks}

The advantages of gradually improving the parameters of the distributional regression model
are manifold, one of which is the ability to apply regularization and variable selection tech-
niques to the evolution of the coefficients. In the gradient boosting setting, some popular tech-
niques include cross-validation \citep{sdr:Hastie+Tibshirani+Friedmann:2009},
stability selection \citep{Meinshausen2010, sdr:Thomas2018GradientBF},
and variable deselection \citep{Stromer2022}. Cross-validation
is a common regularization method used to find an early stopping for the algorithm. 
Stability selection is another method that involves randomly dividing the data into multiple
subsets, boosting the regression model on each subset, and then selecting variables that are
consistently selected across different subsets.
However, both methods have the disadvantage of requiring the model to be calculated multiple times
on different subsets, which can drastically increase computation time. Variable deselection is a variable
selection algorithm that first grows a full model via gradient boosting and then deselects
unimportant variables with low impact on risk reduction. We use these methods in the
simulation in Section~\ref{sec:simulation} as benchmark methods in combination with the
non-cyclical gradient boosting.

For the cyclical updating, cross-validation aims to find an early stopping for each distributional parameter. However, this method can be computationally expensive as a multidimensional grid of stopping iterations needs to be considered, which makes it unsuitable for many (big data) applications. Therefore, we only compare non-cyclical updating with our new methods in the simulation.

One of the main challenges in dealing with complex distributions is the selective saturation of
some distributional parameters that correspond to small gradient vectors $\mathbf{g}_{k^*}$. These
parameters may be selected late in the updating process or not at all, while other parameters may
already be saturated with selected variables.
As a result, achieving a balanced selection of variables for different parameters may prove challenging, potentially resulting in reduced predictive performance and, more importantly, incorrect conclusions regarding influencing factors. Furthermore, the log-likelihood function for complex distributions may have multiple
local maxima, which increases the risk of getting stuck in such maxima due to the vanishing
gradient problem. To address these issues, we propose an adaptation of the general stagewise
regression algorithm \citep{Tibshirani2015} for distributional regression. In the next section,
we provide a detailed description of our approach and discuss its improvements.

\section{Stagewise Boosting} \label{sec:stag}

In this section, we adapt the general stagewise regression presented in \citet{Tibshirani2015}
to the distributional regression setting and present several notable improvements. Please note that
in the following we use non-cyclical updating (see Section~\ref{sec:noncyclic}) throughout.

\subsection{Stagewise Boosting Distributional Regression} \label{sec:sdr}

We start with adapting the forward stagewise algorithm in \citet{Tibshirani2015} for a distributional gradient boosting setting similar to gradient boosting (\ref{eqn:upstep1}) as follows.
Assume all covariates are standardized, we then replace the tentative updates in (\ref{eqn:upstep1}) with,
\begin{equation} \label{diststagstep}
\boldsymbol{\beta}_{k}^{[t]} = \boldsymbol{\beta}_{k}^{[t-1]} +
  \epsilon \cdot \mathrm{sign}\left(({\mathbf{X}_k})_{\cdot j_k^*}^{\top}
  \mathbf{g}_{k}\right) \cdot \mathbf{e}_{j_{k}^*},
\end{equation}
where $\mathbf{g}_k$ is defined in (\ref{eqn:gradvec}). Furthermore, we replace the RSS-criterion with the inner product criterion (IP), meaning the index $j_k^*$ is selected as the maximizing index in absolut value among $1, \dots ,J_k$ for the inner product  $({\mathbf{X}_k})_{\cdot j_k^*}^{\top}\mathbf{g}_{k}$, i.e.,
$$
j_k^* = \underset{{j=1,\dots,J_k}}{\mathrm{argmax}} \, \left|
  ({\mathbf{X}_k})_{\cdot j}^{\top}\mathbf{g}_{k} \right|.
$$

Using the linearity of the predictors $\eta_k$ and the chain rule, one can derive the following equality,
\begin{align*}
({\mathbf{X}_k})_{\cdot j_k^*}^{\top}\mathbf{g}_{k} &= 
\sum_{i = 1}^n ({\mathbf{X}_k})_{i j_k^*} \frac{\partial \, \log \, d_y(y_i; \theta_{i1}^{[t-1]}, \ldots, \theta_{iK}^{[t-1]})}{\partial \eta_k} \\
 &= \sum_{i = 1}^n \frac{\partial \eta_{ik}}{\partial \beta_{ j_k^*}}
 \frac{\partial \, \log \, d_y(y_i; \theta_{i1}^{[t-1]}, \ldots, \theta_{iK}^{[t-1]})}{\partial \eta_k}  =
\frac{\partial \, \ell(\boldsymbol{\beta}^{[t-1]};\mathbf{y},\mathbf{X})}{\partial \beta_{j_k^*k}}.
\end{align*}
This means the inner product $({\mathbf{X}_k})_{\cdot j_k^*}^{\top}
  \mathbf{g}_{k}$ coincides with the derivative of the log-likelihood with respect to $\beta_{j_k^*k}$.


Since we have standardized variables, the correlation $c_{j_kk}$ is given by

\[
c_{j_kk} = \frac{1}{n-1} (\mathbf{X}_{k})_{\cdot j_k}^{\top} \mathbf{g}_{k}.
\]

Furthermore, the residual sum of squares (RSS) can be expressed in terms of the correlation as follows:

\[
\left\| \mathbf{g}_k - (\mathbf{X}_{k})_{\cdot j_k} c_{j_k} \right\|_2^2 = \mathbf{g}_k^{\top} \mathbf{g}_k \left(1 - c_{j_kk}^2 \right).
\]

This shows that the RSS and the inner product (IP) criteria are equivalent.

Thus, the only difference to gradient boosting is that the update step length is constant,
denoted by $\epsilon$.
Since a constant step length is used, the stagewise boosting update rule is not susceptible to
the vanishing gradient problem. However, a disadvantage of (\ref{diststagstep}) is that we lose
the flexibility to take larger update steps when the gradient information would suggest it,
requiring more steps than the usual gradient boosting update (\ref{eqn:upstep1}). In addition,
(\ref{diststagstep}) is unable to improve the coefficients in the convergence phase of the
algorithm, when much smaller updates than $\epsilon$ are required. We fix both problems with further
adjustments to (\ref{diststagstep}), namely with a semi-constant stagewise version similar to the
gradient clipping in \citet{bengio1994}. Moreover, after $\rho\cdot 100$\% (e.g., 80\%) of
iterations, we no longer impose a minimum length of the update step, allowing the algorithm to
converge extremely fast.

Gradient (or partial derivative) clipping is a technique often used when training
deep learning models to prevent gradients from becoming too large or too small. It involves
rescaling the gradients so that their (Euclidean-) length is within a certain range. This is usually done by
setting a maximum (minimum) threshold for the gradients. If the length of a gradient is larger
(smaller) than this value, it is scaled so that it has the same length as the threshold value. 

\paragraph{Semi-constant step length}

More precisely, we substitute the constant $\epsilon$ in (\ref{diststagstep}) with
the semi-constant version $\epsilon_{j^*_kk}$, which depends on the partial derivative of the
normalized log-likelihood $\ell(\boldsymbol{\beta}; \mathbf{y}, \mathbf{X}) / n$ with respect to the $j^*_k\text{-th}$ variable in the $k\text{-th}$
distributional parameter, denoted by $\beta_{j^*_kk}$\footnote{Instead of the sum, we use the
mean to compute the gradients to avoid dependence on the sample size.}.
With this, we arrive at the updating rule
\begin{align*} \label{eq:noncyc}
\partial \ell_{ j^*_k k} &:= \frac{1}{n}\frac{\partial \ell(\boldsymbol{\beta}^{[t-1]} ; \mathbf{y}, \mathbf{X})}{\partial{\beta_{j^*_kk}}}, \\
\epsilon_{j^*_k k} &= \begin{cases}
             \nu\cdot \epsilon   & \text{if } | \partial \ell_{ j^*_k k} | < \nu \cdot \epsilon \text{ and } t <  \rho \cdot T \\
            | \partial \ell_{ j^*_k k} |  & \text{if } | \partial \ell_{ j^*_k k} | < \nu \cdot \epsilon \text{ and } t \geq  \rho\cdot T\\\tag{SC-SDR}
             | \partial \ell_{ j^*_k k} | & \text{if } \nu \cdot \epsilon \leq | \partial \ell_{ j^*_k k} | \leq  \epsilon \\
             \epsilon   & \text{else},
       \end{cases}\\
\boldsymbol{\beta}_{k}^{[t]} &= \boldsymbol{\beta}_{k}^{[t-1]} + \epsilon_{j^*_k k} \cdot \mathrm{sign}\left((\mathbf{X}_{k})_{\cdot j^*_k}^{\top}  \mathbf{g}_{k}\right) \cdot \mathbf{e}_{j^*_k}\\
j_k^* &= \underset{{j=1,\dots,J_k}}{\mathrm{argmax}} \, \left|
  ({\mathbf{X}_k})_{\cdot j}^{\top}\mathbf{g}_{k} \right|, 
\end{align*}

where $\nu$ (e.g., 0.1) is a shrinkage parameter used to define the range of the update. 
Essentially, our update stays between $\nu \cdot \epsilon_{j^*_k k} \cdot \mathrm{sign}\left(({\mathbf{X}_{k}})_{\cdot j^*_k}^{\top}  \mathbf{g}_{k}\right)$ and $\epsilon_{j^*_k k} \cdot \mathrm{sign}\left(({\mathbf{X}_{k}})_{\cdot j^*_k}^{\top}  \mathbf{g}_{k}\right)$ in the first $\rho\cdot 100$\% of the iterations. After that, we omit the minimum threshold and the update stays between $0$ and $\epsilon_{j^*_k k} \cdot \mathrm{sign}\left(({\mathbf{X}_{k}})_{\cdot j^*_k}^{\top}  \mathbf{g}_{k}\right)$.
 As mentioned earlier, this makes it possible for the algorithm to converge properly. The tag of \ref{eq:noncyc} stands for semi-constant updating for stagewise distributional regression.

\paragraph{Illustration}

\begin{figure}[ht!]
\centering
\includegraphics[width=0.9\textwidth]{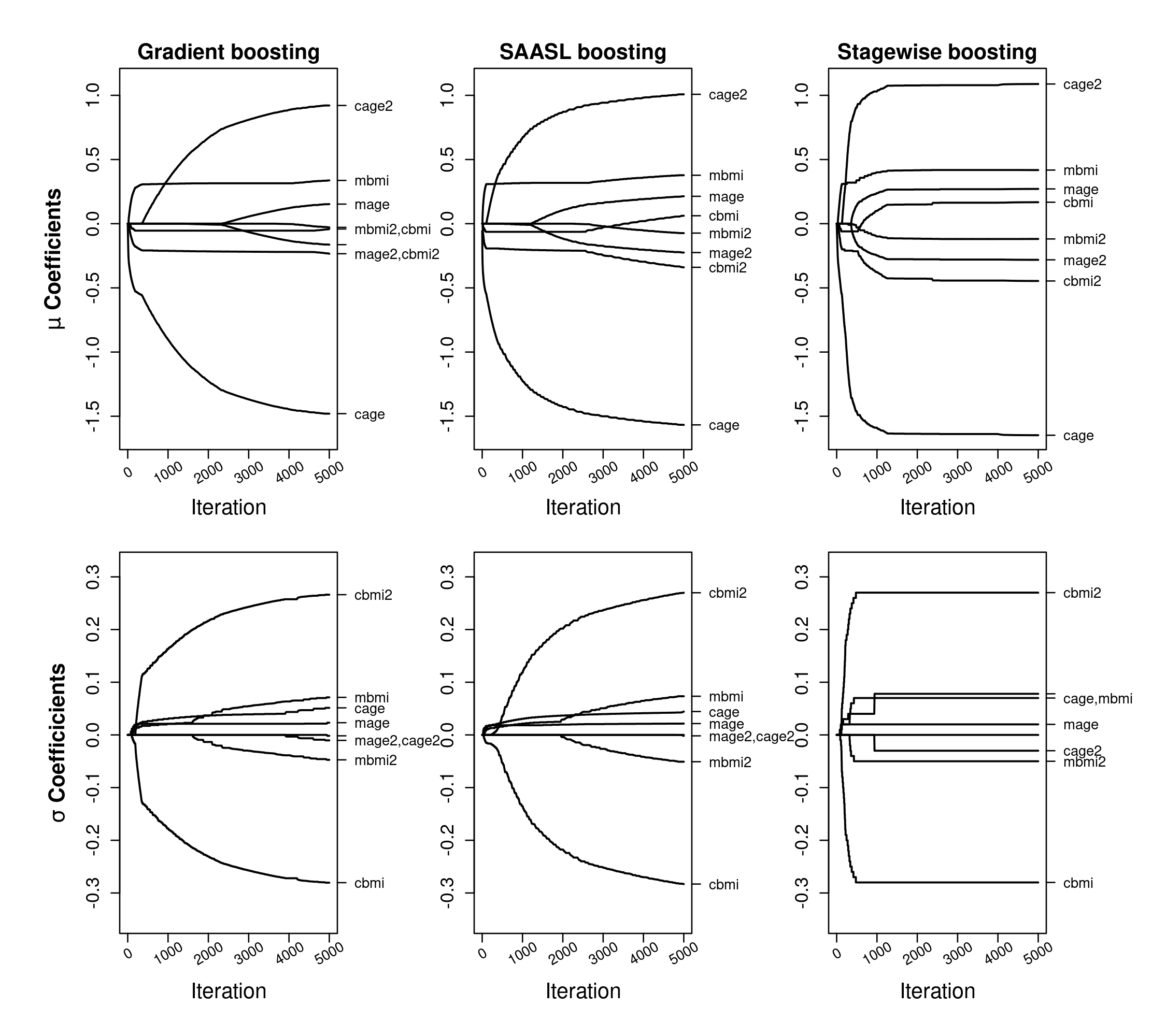}
\caption{\label{fig:stunting} Comparison of coefficient paths obtained from different boosting
  algorithms to illustrate the problem of chronic malnutrition of preschool children in India. 
  The three methods shown are gradient boosting, SAASL boosting, and non-cyclical stagewise 
  boosting with updating rule \ref{eq:noncyc}.}
\end{figure}

We demonstrate the effectiveness of \ref{eq:noncyc} by using the example on chronic undernutrition of preschool children 
taken from \citet{sdr:2102.09248}. In this manuscript, the authors develop a semi-analytical adaptive step-length (SAASL) 
boosting algorithm that aims to address the problem of unequal updating in distributional regression for the normal
location-scale model in the context of non-cyclical gradient boosting. The updating step length therein is 10\% 
of the optimal updating step length with regard to the risk reduction of the respective parameters. We also compare our 
approach to the non-cyclical gradient boosting algorithm as implemented in the \proglang{R} package \pkg{gamboostLSS} 
\citep{sdr:Hofner2021}. This application illustrates modeling chronic undernutrition of preschool children aged 0-35 
months\textemdash denoted by the variable \code{stunting}\textemdash for the same subset used by \citet{sdr:2102.09248} of the data taken from the 1998-99 Indian Demographic and Health Survey \citep{DHS1998-99IND}. Similar to \citet{sdr:2102.09248} a normal model $\texttt{NO}(\mu = \eta_{\mu}, \log(\sigma) = \eta_{\sigma})$ is used,
where predictors $\eta_{k}$, $k \in \{\mu, \sigma\}$, depend on a quadratic polynomial of each
of the variables $\texttt{cbmi}$ (BMI of the child), $\texttt{cAge}$ (age of the child),
$\texttt{mBMI}$ (BMI of the mother), and $\texttt{mAge}$ (age of the mother) and are specified as follows:
\begin{eqnarray*}
\eta_k &=& \beta_{0k} + \beta_{1k} \cdot \texttt{cBMI} + \beta_{2k} \cdot \texttt{caAge} +
              \beta_{3k} \cdot \texttt{mBMI} + \beta_{4k} \cdot \texttt{mAge}+ \\
	         && \beta_{5k} \cdot \texttt{cBMI2} + \beta_{6k} \cdot \texttt{cAge2} + \beta_{7k} \cdot \texttt{mBMI2} +
              \beta_{8k} \cdot \texttt{mAge2}.
\end{eqnarray*}
All variables are standardized and the added \code{2} at the end of the variable names denotes
the squared effects. In Figure~\ref{fig:stunting}, the corresponding coefficient paths are shown.
We observe that both the non-cyclical gradient boosting and SAASL-boosting algorithm fail to
converge within $5000$ iterations, while the \ref{eq:noncyc} algorithm converges after
$\approx 1300$ iterations. This is due to the small updating steps that depend on the small
gradients in the first two methods and the clipped updating steps in the latter ($\nu = 0.1, \epsilon = 0.01, \rho = 0.8$).
The results demonstrate that stagewise boosting distributional regression with the updating 
rule \ref{eq:noncyc} can improve the convergence speed and accuracy compared to other methods.

Why is this important? In a cross-validation scenario, consider the task of identifying relevant variables. If the boosting algorithm fails to converge, it may result in a high number of false negatives. While the user might attempt to address this by increasing the number of iterations, our experience suggests that doing so can render estimation infeasible. Additionally, users may not always be aware of this potential issue.

\subsection{Best-Subset Updating}
 
In the non-cyclical updating scheme, only one distributional parameter is updated per 
iteration. In contrast, the cyclical updating scheme updates each parameter in turn at every 
iteration, requiring the selection of optimum stopping iterations for each parameter through a 
grid search. While the latter approach is often computationally infeasible, e.g., for 
stability selection or big data problems, it still offers some advantages in terms of variable 
selection. This is because parameter-wise stopping iterations, or updates, can be generated 
for parameters even when encountering issues like vanishing small gradients, provided that a 
sufficient number of iterations are specified. To leverage the advantages of both approaches, 
we propose a novel updating variant called \emph{best-subset updating}. This approach is 
particularly useful for estimating models with complex distributions, such as \code{ZANBI}, 
where certain distributional parameters may be dominated by others using non-cyclical 
updating, making it impossible to select a balanced set of variables. Best-subset updating 
addresses this issue by considering all possible combinations of distributional parameters for 
updating, resulting in superior estimation of all distributional parameters. This approach 
mirrors the computationally demanding grid search employed in cyclic boosting algorithms, but 
instead of tracking a grid of parameter iterations, only one iteration path is required. This 
reduces the early stopping search to a single dimension, which makes the best-subset updating 
very useful for big data problems. For example, this can be seen in the simulation in 
Section~\ref{sec:simulation}, where we compare the true positive rate of the three parameter 
distributions in Figure~\ref{fig:res100ZANBI}.

The algorithm works as follows: For every non-empty subset $S \subset \{1,\dots, K\}$,
we combine the partial derivatives corresponding to $S$ into a gradient vector
$\nabla \mathbf{L}_S = \left(\partial \ell_{ j^*_s s}\right)_{s \in S}$ which will be used to 
define the updating step. These partial derivatives represent a weighting of the different 
variables with regard to their relative importance. To avoid an exploding or a vanishing 
gradient, we rescale the gradient vector $\nabla \mathbf{L}_S$ if its Euclidean length exceeds 
a certain value $\epsilon$.
Additionally, if any individual partial derivative $\partial \ell_{ j^*_{s} s}$ fails to overcome a
minimum threshold $\nu\cdot \epsilon$ (e.g., $\nu = 0.1$ and $\epsilon = 0.01$) in the first
$\rho\cdot 100\%$ of the iterations (e.g., $\rho = 0.8$), we rescale them to this threshold value.
The scaling factor controlling the maximum threshold is defined as
\begin{align*} 
\epsilon_{S} &= \begin{cases}
  1  & \text{if } \| \nabla \mathbf{L}_S \|_2 < \epsilon \\
  \frac{\epsilon}{\| \nabla \mathbf{L}_S \|_2}  & \text{else}.
\end{cases}
\end{align*}
Then, clipping from below for the individual derivatives ensures that the updating steps do
not vanish in the first $\rho\cdot 100$\% of the iterations and the final updating step is
\begin{align*} 
\tag{SC-BS-SDR}\label{eq:bestsub}
\epsilon_{Ss} &= \begin{cases}
             \nu\cdot \epsilon   & \text{if }  \epsilon_{S}\cdot |\partial \ell_{ j^*_s s} | < \nu \cdot \epsilon \text{ and } t <  \rho \cdot T \\
            \epsilon_{S}\cdot |\partial \ell_{ j^*_s s}|  & \text{if }  \epsilon_{S}\cdot |\partial \ell_{ j^*_s s}| < \nu \cdot \epsilon \text{ and } t \geq  \rho\cdot T\\
             \epsilon_{S}\cdot |\partial \ell_{ j^*_s s}| & \text{if } \nu \cdot \epsilon \leq  \epsilon_{S}\cdot |\partial \ell_{ j^*_s s}| \leq  \epsilon \\
             \epsilon   & \text{else},
       \end{cases}\\
\boldsymbol{\beta}_{s}^{[t]} &= \boldsymbol{\beta}_{s}^{[t-1]} + \epsilon_{Ss} \cdot \mathrm{sign}\left(({\mathbf{X}_{s}})_{\cdot j^*_{s}}^{\top}  \mathbf{g}_{s}\right) \cdot \mathbf{e}_{j^*_{s}}\\
j_s^* &= \underset{{j=1,\dots,J_s}}{\mathrm{argmax}} \, \left|
  ({\mathbf{X}_s})_{\cdot j}^{\top}\mathbf{g}_{s} \right|,  
\end{align*}
for $s \in S$. Similar to \ref{eq:noncyc}, after $\rho\cdot 100$\% of the iterations, the minimal
length of the updating step $\epsilon_{Ss}$ is no longer imposed. The subset $S^*$ which maximizes
the log-likelihood $\ell(\cdot)$ is choosen for an update in iteration $t$. The tag of \ref{eq:bestsub} stands for semi-constant, best subset updating for stagewise distributional regression.

To prevent overfitting, early stopping methods need to be applied to both \ref{eq:noncyc} and
\ref{eq:bestsub}, as both algorithms gradually build up their coefficients. In the following,
we introduce a novel variable selection method, correlation filtering, that takes advantage of the correlation of
the data with the gradient vectors in each iteration. This variable selection algorithm can
be used in combination with both \ref{eq:noncyc} and \ref{eq:bestsub}.

\subsection{Early Stopping} \label{sec:earlystopping}

Optimal stopping iterations in boosting settings are typically determined through costly cross-validation, which can be impractical in many scenarios. To address this, we propose using the Bayesian information criterion (BIC): 
$$
\mathrm{BIC}^{[t]} = -2\cdot \ell(\boldsymbol{\beta}^{[t]} ; \mathbf{y}, \mathbf{X})  + \log(n)\cdot \mathrm{df}^{[t]},
$$
where $\mathrm{df}^{[t]}$ is the number of non-zero parameters at iteration $t$. This approach 
usually leads to parsimonious models, since counting the number of non-zero coefficients is 
more restrictive compared to estimating the degrees of freedom using the hat matrix (which is 
very computationally intensive). \citet{Zou:2007} investigate the usefulness of this approach 
for LASSO. We adopt this approach for stagewise boosting distributional regression.

\subsection{Correlation Filtering} \label{sec:cfilter}

Instead of computing the inner product of the variables with the gradient vector ${(\mathbf{X}_{k}})_{\cdot j_k}^{\top}  \mathbf{g}_{k}$ for each distributional parameter used to determine the candidate variables to be updated, one can determine the variable to be updated using the correlation between the gradient vector and the variables if all variables are standardized. The latter variant has the advantage that it can be regularized by adding a threshold value $\kappa$ (e.g. $\kappa = 0.15$) to filter out variables with an absolut correlation value $\left|c_{j_kk}\right| \leq \kappa$. Only the remaining variables with $\left|c_{j_kk}\right|>\kappa$ are considered for a possible update. In the case that no variable remains in a distributional parameter with a sufficiently large correlation $c_{j_kk}$ with $\mathbf{g}_{k}$, then no candidate variable is provided. In the extreme case where no candidate variable is provided across all distributional parameters, no update is carried out. Typically, in the early iterations of the updating, the correlation between the covariates and $\mathbf{g}_{k}$ are large and with continuing updates the correlated variables get absorbed into the model, leading to a decline of the correlation values until they all no longer overcome the minimum requiremend $\kappa$ for an update. At this stage the updating stops and an implicit early stopping is found.
The correlation filtering (CF) in combination with \ref{eq:bestsub} is presented in Algorithm~\ref{alg:alg1} in Section~\ref{sec:EffEst}.

Please note that CF only serves as a variable selection step. After selecting the variables, 
their coefficients are boosted in a refitting step until convergence.
This process resembles the deselection algorithm presented by \citet{Stromer2022}, where a 
complete model, incorporating all variables, is grown using boosting\footnote{utilizing non-
cyclical gradient boosting in the distributional regression setting}. Subsequently, 
insignificant variables with minimal impact on risk reduction are deselected, and the model is 
re-estimated using only the selected variables (algorithm VarDes in the following).

\begin{figure}[ht!]
\centering
\includegraphics[width=1\textwidth]{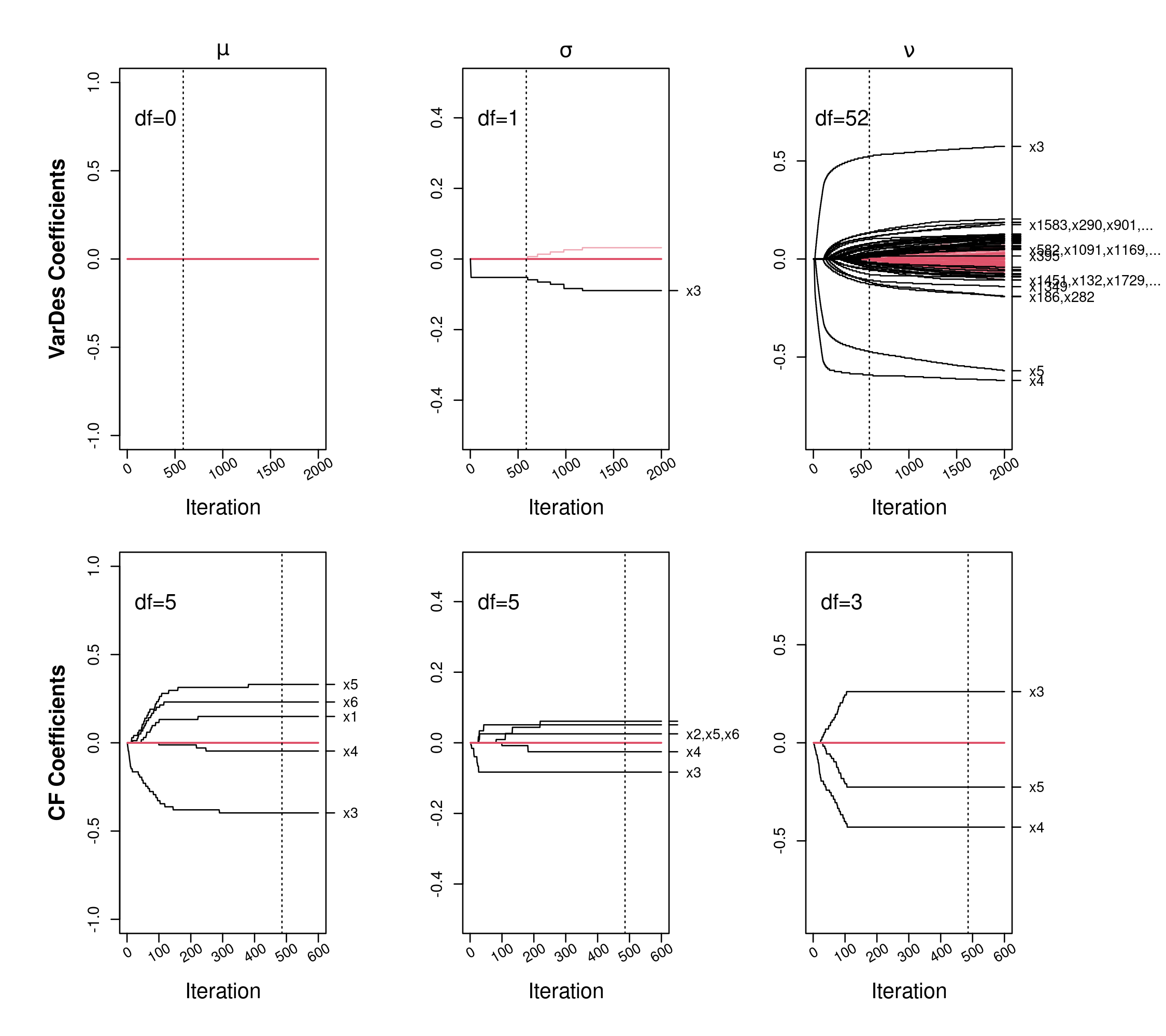}
\caption{\label{fig:compZANBI} Comparison of variable selection steps of VarDes (top) vs.\ CF
  (bottom). \code{ZANBI} distribution
  with 1000 observations and 2000 noise variables for parameters $\mu$, $\sigma$ and $\nu$. The true non-noise variables for $\mu$ are
  $x_1, x_3, x_5, x_6$, for $\sigma$ are $x_2, x_4, x_5$ and for
  $\nu$ are $x_3, x_4, x_5$. All variables are correlated according to the correlation
  setting in the simulation Section~\ref{sec:simulation} ($\rho = 0.7$). The vertical-gray-dotted line represents
  the early stopping induced by 10-fold cross validation for the VarDes and BIC by CF. The displayed variable names are the
  final set of selected variables for the respective predictors.}
\end{figure}

To exemplify the usefulness of CF, a comparison between VarDes and CF with best-subset selection is made in Figure~\ref{fig:compZANBI}.
This case shows a high $p$ low $n$ setting ($p>n$, we simulate data according to the data generating process presented in the simulation in Table~\ref{tab:dists}). This dataset has 1000 observations and 2000 correlated noise variables for each predictor.
While the VarDes algorithm effectively identifies true positive variables for parameter $\nu$, 
it fails to select any variables for parameter $\mu$, and only selects one false variable for $\sigma$. The absence of true variable selections for parameter $\mu$ and $\sigma$ is 
attributed to small gradients concerning this predictors, which make all variables appear 
insignificant for the gradient boosting algorithm. Conversely, CF identifies all 
true variables and demonstrates fewer falsely selected variables. Overall, our correlation filtering selects 10 out of 10 true effects and 3 false effects. Variable deselection identified 3 out of 10 true effects and selects 50 false effects.

\subsubsection{Choice of the Correlation Threshold - Hypothesis Testing Framework}

As the sample size and number of variables vary, the appropriateness of setting $\kappa$ to an arbitrary value becomes questionable. A smaller value might be necessary as the sample size increases, while a larger value may be warranted with a greater number of variables. Therefore, $\kappa$ should be adjusted accordingly to better reflect the nuances of each situation. Therefore we propose to set $\kappa$ as the critical value in the following hypothesis testing framework. Assume we are dealing with the variables $(\mathbf{X}_{k})_{\cdot 1}, \dots, (\mathbf{X}_{k})_{\cdot J_k}$ for a given parameter $\theta_k$. Furthermore assume all variables and the gradient vector $\mathbf{g}_{k}$ are standardized. 

The sample version of the Pearson correlation coefficient is given by:
\begin{align*}
c_{j_kk} = \frac{1}{n-1}({\mathbf{X}_{k}})_{\cdot j_{k}}^{\top}  \mathbf{g}_{k} = \frac{\sqrt{n}}{n-1} \frac{1}{\sqrt{n}}({\mathbf{X}_{k}})_{\cdot j_{k}}^{\top}  \mathbf{g}_{k} 
\end{align*}   
and let $C_{j_kk}$ denote the population version of the Pearson correlation coefficient.\\
The hypothesis pair:
\begin{align*}
H_0\text{: } R_{j^*_kk}=&\mathrm{max}\left\{\lvert C_{1k} \rvert,\dots ,\lvert C_{J_kk}\rvert \right\}=0,
\text{ where we assume independence of } ({\mathbf{X}_{k}})_{\cdot j_{k}} \text{ and }  \mathbf{g}_{k}.\\
H_1\text{: } R_{j^*_kk}=&\mathrm{max}\left\{\lvert C_{1k} \rvert,\dots ,\lvert C_{J_kk}\rvert \right\}\neq0.
\end{align*}

The standardization and independence under $H_0$ of ${(\mathbf{X}_{k}})_{\cdot j_{k}}$ and $\mathbf{g}_{k}$ implies the random variable corresponding to the inner product ${(\mathbf{X}_{k}})_{\cdot j_{k}}^{\top}  \mathbf{g}_{k}$ is a sum of $n$ products, where each product has mean zero and variance one, and thus $\frac{1}{\sqrt{n}}({\mathbf{X}_{k}})_{\cdot j_{k}}^{\top}  \mathbf{g}_{k} \stackrel{asy.}{\rightarrow} \mathcal{N}\left(\mu = 0,\sigma^2 = 1 \right)$ according to the central limit theorem.
This implies $c_{j_kk}$ follows asymptotically a normal distribution $\mathcal{N}\left(\mu = 0,\sigma^2 = \frac{n}{(n-1)^2} \right)$. 

We take a threshold value which corresponds to a significance level $\alpha$, i.e.,
a threshold value satisfying
\begin{align*}
1-\alpha = \mathbb{P}\left(\mathrm{max}\left\{\lvert C_{1k} \rvert,\dots \lvert C_{J_kk} \rvert)\right\} \leq \kappa | H_0\right).
\end{align*}   

The independence stated in $H_0$ implies the pairwise independence of $C_{1k},\dots, C_{J_kk}$, thus
$$
1-\alpha = \mathbb{P}\left( -\kappa \leq C_{1k} \leq \kappa | H_0\right)^{J_k}.
$$
As the asymptotic distribution for $C_{1k}$ is a normal distribution, we can easily compute the corresponding quantiles and thus the threshold value $\kappa$ with help of the inverse of the cumulative distribution function of the standard normal distribution $\Phi^{-1}(p)$:
$$
\kappa = \Phi^{-1}\left( \frac{1+( 1-\alpha)^{\frac{1}{J_k}}}{2}\right)\cdot \frac{\sqrt{n}}{n-1}. 
$$

The independence assumption might appear stringent but at the later stages of the updating, when a lot of the dependence is absorbed into the model, the remaining dependence should be negligible and before that, the correlation is typically high. We use $\alpha = 0.05$ as the default setting.
As the simulations show, this approximation works quite well for the correlation filtering. 

\subsubsection{Choice of the Correlation Threshold - Threshold Descent Algorithm}

\begin{figure}[ht!]
\centering
\includegraphics[width=1\textwidth]{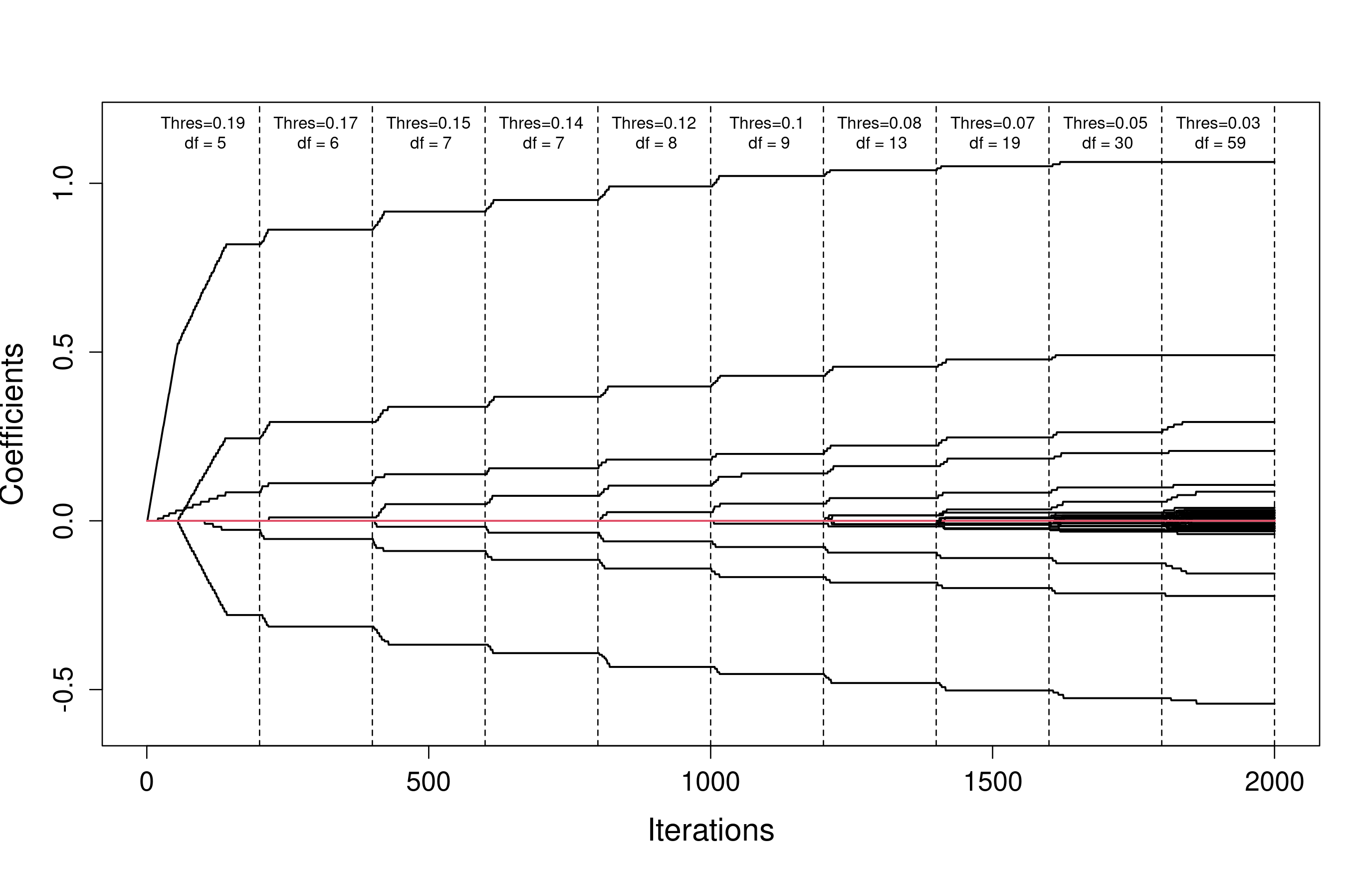}
\caption{\label{fig:thresdesc} Illustration of threshold descent on simulated \code{NO} data. Displayed are the coefficients corresponding to parameter $\mu$. Starting with a high CF value $\kappa = 0.19$, the algorithm boosts for 200 iterations thereby selecting a few variables. Every 200 iterations $\kappa$ gets lowered by 0.02, thereby relaxing the correlation filtering threshold. This in turn allows for more variables to enter the model the lower the threshold descends. }
\end{figure}

An alternative approach to determining the value of $\kappa$ for the CF method involves employing a grid search framework.
Threshold descent works by starting with a high $\kappa$ value and by boosting the model until the coefficients no longer overcome the initial threshold. This gives us the first set of coefficients, which are then refitted in a later step. Then, the value $\kappa$ is lowered and continuing the boosting, additional variables may enter the model. The second set of coefficients is saved and, once again, the threshold is lowered. This process is continued until the threshold $\kappa = 0$.  
These are the selection steps. For the final model, the different sets of the selection process are then refitted until convergence. The final model is then determined by the BIC. The threshold descent algorithm is illustrated in Figure~\ref{fig:thresdesc}.   

As the observation size and the number of variables grow ($n$ and $p$), the computational burden as well as the need for fast algorithms increases. Therefore we further propose a batched version of the CF, \ref{eq:noncyc} and \ref{eq:bestsub}, similar to stochastic gradient descend. 

\subsection{Efficient Estimation for Large-Scale Data} \label{sec:EffEst}

In this subsection, we adopt a stochastic approximation approach for faster memory efficient
computation of inner products and updating steps, following the methodology proposed by
\citet{umlauf2023}. Specifically, we compute the inner products
$(\mathbf{X}{k})_{[\mathbf{i}_t] j^*_k}^{\top} \mathbf{g}_{[\mathbf{i}_t] k}$ using only a randomly
selected batch $\mathbf{i}_t \subset \{1,\dots,n\}$ of size $bs$ at each iteration.
These inner products are used to identify the candidate variables for updating for each predictor.

For the updating step of \ref{eq:noncyc} we replace the normalized derivatives with their batchwise
counterpart based only on the data $\mathbf{i}_t$:
\begin{align*}
\label{eq:bwnoncyc}
\partial \hat{\ell}_{ j_k^* k} &= \frac{1}{bs}\frac{\partial \ell(\boldsymbol{\beta}^{[t-1]} ; \mathbf{y}_{[\mathbf{i}_t]}, \mathbf{X}_{[\mathbf{i}_t]})}{\partial{\beta_{j_k^* k}}}, \\
\hat{\epsilon}_{ j^*_k k } &= \begin{cases}
             \nu\cdot \epsilon   & \text{if } | \partial \hat{\ell}_{ j_k^* k } | < \nu \cdot \epsilon \text{ and } t <  \rho \cdot T \\
            | \partial \hat{\ell}_{ j_k^* k} |  & \text{if } | \partial \hat{\ell}_{ j_k^* k} | < \nu \cdot \epsilon \text{ and } t \geq  \rho\cdot T\\\tag{SC-BW-SDR }
             | \partial \hat{\ell}_{ j_k^* k } | & \text{if } \nu \cdot \epsilon \leq | \partial \hat{\ell}_{ j_k^* k } | \leq  \epsilon \\
             \epsilon   & \text{else},
       \end{cases}\\
\boldsymbol{\beta}_{k}^{[t]} &= \boldsymbol{\beta}_{k}^{[t-1]} + \hat{\epsilon}_{ j_k^* k} \cdot \mathrm{sign}\left((\mathbf{X}_{k})_{[\mathbf{i}_t] j_k^*}^{\top}  \mathbf{g}_{[\mathbf{i}_t] k}\right) \cdot \mathbf{e}_{j_k^*}\\
j_k^* &= \underset{{j=1,\dots,J_k}}{\mathrm{argmax}} \, \left|
  ({\mathbf{X}_k})_{[\mathbf{i}_t] j}^{\top}\mathbf{g}_{k} \right|.
\end{align*}

The batchwise variant of \ref{eq:bestsub} is analogous to \ref{eq:bwnoncyc}. Both batchwise
variants perform their preliminary variable selection on a subset $\mathbf{i}_t$ of the data.
The decision which distribution parameter or combination of parameters to update, if any,
is made on the next batch $\mathbf{i}_{t+1}$. This leads to additional stability as the
second step is performed on data that is quasi out-of-sample. An additional advantage is also
that by randomization the risk of getting stuck in local optima can be substantially minimized when using more complex distributions. The exact algorithm is shown in Algorithm~\ref{alg:alg1}.
Specifying a threshold of zero yields the correlation filter-free algorithm, and replacing all
possible subsets with only the singleton subsets in the best-subset selection yields the
non-cyclic update.

For batchwise correlation filtering, a threshold derived from the assumption\newline
$C_j \stackrel{a}{\sim} \mathcal{N}\left(\mu = 0,\sigma^2 = \frac{bs}{(bs-1)^2} \right)$ is used.

\begin{algorithm}[p!]
\spacingset{0.9}\footnotesize
\renewcommand\thealgorithm{SBDR}
\caption{:\label{alg:alg1} Stagewise Boosting Distributional Regression.}
\begin{algorithmic}
\State Set $\epsilon > 0$, $T > 1$ and $\kappa,\rho,\nu  \geq 0$,
  batch indices $B = (\mathbf{i}_1,\dots, \mathbf{i}_T)$ of size $bs = |\mathbf{i}_1|$.
\State Initialize $\boldsymbol{\beta}^{[1]}$. Intercept-coefficients are initialized at corresponding maximum likelihood estimates, all other coefficients at zero. Set $\boldsymbol{\beta}^{[2]} = \boldsymbol{\beta}^{[1]}$.
\For{t from 2 to $T$}
	\State Set current batch $\mathbf{i} = \mathbf{i}_t$ and next batch
    $\tilde{\mathbf{i}} = \mathbf{i}_{t+1}$ for $t < T$, if $t = T$,
    then $\tilde{\mathbf{i}} = \mathbf{i}_{1}$.
	\For{k from 1 to K }	
		\State $\boldsymbol{\eta}_{[\mathbf{i}] k} \leftarrow \mathbf{X}_{[\mathbf{i}]k} \boldsymbol{\beta}_{k}^{[t-1]}$ and
		  $\boldsymbol{\eta}_{[\tilde{\mathbf{i}}] k} \leftarrow \mathbf{X}_{[\tilde{\mathbf{i}}]k} \boldsymbol{\beta}_{k}^{[t-1]}$.
	\EndFor
	\State Compute $\texttt{logLik}_{[\mathbf{i}]} \leftarrow \ell(\mathbf{y}_{[\mathbf{i}]},
    \boldsymbol{\eta}_{[\mathbf{i}] 1}, \dots, \boldsymbol{\eta}_{[\mathbf{i}] K})$ and
	  $\texttt{logLik}_{[\tilde{\mathbf{i}}]}^{\texttt{old}} \leftarrow \ell(\mathbf{y}_{[\tilde{\mathbf{i}}]},
    \boldsymbol{\eta}_{[\tilde{\mathbf{i}}] 1}, \dots, \boldsymbol{\eta}_{[\tilde{\mathbf{i}}] K})$.
	\For{k from 1 to K }	
		\State Update intercept $\beta_{0k}^{[t]}$ with clipped gradient update with $\frac{1}{bs}\frac{\partial \ell(\mathbf{y}_{[\mathbf{i}]}, \boldsymbol{\eta}_{[\mathbf{i}] 1} , \dots, \boldsymbol{\eta}_{[\mathbf{i}] K} )}{\partial{\beta_{0k}}}$. 
	\EndFor

	\For{k from 1 to K}	
	\State $\mathbf{g}_k \leftarrow \left(\frac{\partial \log d_y(y_i,\boldsymbol{\eta}_{i 1} , \dots, \boldsymbol{\eta}_{i K})}{\partial{\eta_k}}\right)_{i\in [\mathbf{i}]}   $
	\State $j_k^* \leftarrow$ Determine non-intercept column of $\mathbf{X}_{[\mathbf{i}]k}$ with highest correlation with $\mathbf{g}_k$ in absolut value.
	\State $c_{j^*_kk} \leftarrow$ Corresponding correlation in absolute value. 
	\If{$c_{j^*_kk} > \kappa$}
	\State $\partial \hat{\ell}_{ j_k^* k} \leftarrow \frac{1}{bs}\frac{\partial \ell(\mathbf{y}_{[\mathbf{i}]}, \boldsymbol{\eta}_{[\mathbf{i}] 1} , \dots, \boldsymbol{\eta}_{[\mathbf{i}] K} )}{\partial{\beta_{j_k^* k}}}$
	\Else
	\State $\partial \hat{\ell}_{j_k^* k} \leftarrow 0 $
	\EndIf
	\EndFor

\State Best-subset selection:
\State $\boldsymbol{\beta}_1^{\mathrm{best}},\dots,\boldsymbol{\beta}_K^{\mathrm{best}} \leftarrow \boldsymbol{\beta}_1^{[t-1]},\dots,\boldsymbol{\beta}_K^{[t-1]}$
\State $\boldsymbol{\beta}_1^{\mathrm{temp}},\dots,\boldsymbol{\beta}_K^{\mathrm{temp}} \leftarrow \boldsymbol{\beta}_1^{[t]},\dots,\boldsymbol{\beta}_K^{[t]}$
\For{non-empty subset   $S \subset \{1,\dots,K \}$}
\State $\nabla \hat{\mathbf{L}}_S \leftarrow \left(\partial \hat{\ell}_{ j^*_{s} s}\right)^{\top}_{s \in S}$
\If{$\left\|\nabla \hat{\mathbf{L}}_S\right\|_2  > \epsilon$}
\State Clip from above
\For{s in S}
\State $\partial \hat{\ell}_{ j^*_{s} s} \leftarrow \frac{\partial \hat{\ell}_{ j^*_{s} s}}{\left\|\nabla \hat{\mathbf{L}}_S\right\|_2} $
\EndFor
\EndIf
\If{$t < \rho\cdot T$}
\State Clip from below
\For{s in S}
\State $\partial \hat{\ell}_{ j^*_{s} s} = \mathrm{sign}(\partial \hat{\ell}_{ j^*_{s} s})\cdot \mathrm{max}( \nu \cdot \epsilon ,|\partial \hat{\ell}_{ j^*_{s} s}|)$
\EndFor
\EndIf
\For{s in S} 
\State Update coefficients
\State $\boldsymbol{\beta}_{s}^{[t]}  \leftarrow \boldsymbol{\beta}_{s}^{[t]} + \partial \hat{\ell}_{ j^*_{s} s} \cdot \mathbf{e}_{j_s^*}$ 
\EndFor
\For{k in K} 
\State $\boldsymbol{\eta}_{[\tilde{\mathbf{i}}] k} \leftarrow \mathbf{x}_{[\tilde{\mathbf{i}}]k} \boldsymbol{\beta}_{k}^{[t]}$\EndFor
\State $\texttt{logLik}_{[\tilde{\mathbf{i}}]}^{\texttt{new}} \leftarrow \ell(\mathbf{y}_{[\tilde{\mathbf{i}}]},
    \boldsymbol{\eta}_{[\tilde{\mathbf{i}}] 1}, \dots, \boldsymbol{\eta}_{[\tilde{\mathbf{i}}] K})$
\If{$\texttt{logLik}_{[\tilde{\mathbf{i}}]}^{\texttt{new}} > \texttt{logLik}_{[\tilde{\mathbf{i}}]}^{\texttt{old}}$}
\State If oos-likelihood improves, keep new coefficients.
\State $\boldsymbol{\beta}_1^{\mathrm{best}},\dots,\boldsymbol{\beta}_K^{\mathrm{best}} \leftarrow \boldsymbol{\beta}_1^{[t]},\dots,\boldsymbol{\beta}_K^{[t]}$
\State $\texttt{logLik}_{[\tilde{\mathbf{i}}]}^{\texttt{old}} \leftarrow \texttt{logLik}_{[\tilde{\mathbf{i}}]}^{\texttt{new}}$
\EndIf
\State Reset coefficients
\State $\boldsymbol{\beta}_1^{[t]},\dots,\boldsymbol{\beta}_K^{[t]} \leftarrow \boldsymbol{\beta}_1^{\mathrm{temp}},\dots,\boldsymbol{\beta}_K^{\mathrm{temp}} $
\EndFor
\State Keep best coefficients
\State $\boldsymbol{\beta}_1^{[t]},\dots,\boldsymbol{\beta}_K^{[t]} \leftarrow \boldsymbol{\beta}_1^{\mathrm{best}},\dots,\boldsymbol{\beta}_K^{\mathrm{best}} $
\State $\boldsymbol{\beta}_1^{[t+1]},\dots,\boldsymbol{\beta}_K^{[t+1]} \leftarrow \boldsymbol{\beta}_1^{\mathrm{best}},\dots,\boldsymbol{\beta}_K^{\mathrm{best}} $
\EndFor
\State \Return best of $\left(\boldsymbol{\beta}_1^{[t]},\dots,\boldsymbol{\beta}_K^{[t]}\right)_{t=1,\dots,T}$ according to the BIC.
\end{algorithmic}
\end{algorithm}

\subsection{Software Implementation}

Our stagewise distributional regression methods are implemented in the \proglang{R} \citep{sdr:R} package \pkg{stagewise}~(\url{https://github.com/MattWet/stagewise}). The package supports all distribution families of the \proglang{R} package \pkg{gamlss.dist} \citep{sdr:gamlss.dist}. Examples with code on how to fit the models using the proposed Algorithm~\ref{alg:alg1} are provided in the help pages of the package. In addition, the scripts for the following simulation study and the lightning application are provided. 

\section{Simulation} \label{sec:simulation}

We present a comprehensive evaluation of our proposed methods compared to other benchmarks.
Our focus is on assessing the performance using several metrics, including the root
mean squared error of the additive predictors (RMSE), a (continuous) ranked probability score
((C)RPS; \citealp{sdr:Gneiting+Raftery:2007}), the number of true positives (TP) and false
positives (FP) in the variable selection process, and the elapsed computation time. To accomplish
this, we employ three distinct distributions (\code{NO}, \code{GA} and \code{ZANBI}),
each with varying numbers of additional noise variables ($30$ and $100$), different pairwise
correlations ($0$ and $0.7$) between the variables, and various sample sizes ($500$ up to $1000000$)
to facilitate a thorough comparison of the methods.

\subsection{Methods}

Please note that all of the boosting-type benchmark methods used in our simulation employ
non-cyclical updates (see Section~\ref{sec:noncyclic}). Specifically, we consider the following
benchmark methods:

\begin{itemize}
\item \textit{True Model} (\code{TM}):
Here the true model is known and thus no non-informative variables are incorporated in the model. The model is estimated with the 
standard backfitting method implemented in the \proglang{R} package
\pkg{bamlss}~\citep{sdr:Umlauf+Klein+Simon+Zeileis:2021, sdr:Umlauf+Klein+Zeileis:2016}.

\item \textit{Gradient Boosting} (\code{GB}):
We use the non-cyclical gradient boosting version in
  \citet{sdr:Thomas2018GradientBF}. The optimal stopping iteration (\textit{mstop}) is selected
  by ten-fold cross-validation. The non-cyclical gradient boosting algorithm is implemented
  in the \proglang{R} package \pkg{gamboostLSS} \citep{sdr:Hofner2021}.

\item \textit{Stability Selection} (\code{StabSel}):
Our third benchmark method is the non-cyclical stability selection method
\citep{Meinshausen2010, sdr:Thomas2018GradientBF}, which is also implemented in the
\proglang{R} package \pkg{bamlss}. This method is based on calculcating the variable selection frequencies on 100 resampled datasets for the \code{NO} and \code{GA}. To make the \code{ZANBI} computational feasable, we incorporate only 10 resampled datasets, which still takes around 250 minutes in a challenging \code{ZANBI} setting with 10000 observations for a single replication. For comparision a similar \code{NO} setting with 100 resampled datasets takes around 80 minutes and an increase from 10 to 100 resampled datasets increases the computation time around 5 to 10-fold.

\item \textit{Variable Deselection} (\code{VarDes}):
First a full model via \textit{Gradient Boosting (GB above)} is estimated and then variables wich contribute less than 1\% to the risk reduction up until the \textit{mstop} are deselected. After that the model with the selected variables is refitted with \textit{mstop} iterations \citep{Stromer2022}.

\item \textit{Semi-Analytical Adaptive Step-Length} (\code{SAASL}):
This method is only used for the normal location-scale settings as it is not implemented
for other distributions \citep{sdr:2102.09248}.
Early stopping is determined by ten-fold cross-validation. 
\end{itemize}

As we propose several adaptations of the general stagewise regression algorithm\:\textendash\:noncyclic
stagewise distributional regression, best-subset stagewise distributional regression, a new variable
selection method, correlation filtering, and a batched update expansion of them\:\textendash\:we investigate the performance measures described at the beginning of this section for all
possible combinations. Each method is using the BIC to determine the optimal stopping iteration \emph{mstop}. This is the first step,
the variable selection step. The degrees of freedom for the BIC penalty are chosen strictly as the
number of non-zero coefficients. In the batchwise variants, we use a moving average over
the computed BIC scores of the updates. In the second step, we recompute the model using only
the non-zero coefficients until convergence and without correlation filtering. Specifically,
we evaluate the performance of the following variants of the proposed algorithm:
\begin{itemize}
\item \code{Standard}:
Non-cyclical updating with BIC. We use Algorithm~\ref{alg:alg1} with $\kappa = 0$ and
the restriction that only subsets of the distributional parameters with only a single element
are allowed in the best-subset search. This turns this algorithm effectively into the non-cyclical
updating version where the correlation filtering is turned off. Furthermore the batchsize is set
equal to the number of observations, i.e., the updating steps are selected with full batchsize. 

\item \code{BS}:
  Best-subset updating with BIC. We use Algorithm~\ref{alg:alg1} with $\kappa = 0$ and
  the batchsize set equal to the number of observations, i.e., the updating steps are
  selected with full batchsize. 

\item \code{CF}:
  Non-cyclical updating and correlation filtering with BIC. We use Algorithm~\ref{alg:alg1}
  with the restriction that only subsets of the distributional parameters with only a single
  element are allowed in the best-subset search. Furthermore the batchsize is set equal to
  the number of observations, i.e., the updating steps are selected with full batchsize. 

\item \code{BS+CF}:
  Best-subset Updating and Correlation Filtering with BIC. We use Algorithm~\ref{alg:alg1} and the
  batchsize is set equal to the number of observations, i.e., the updating steps are
  selected with full batchsize. 

\item \code{BW}:
  Batchwise non-cyclical updating with BIC. We use Algorithm~\ref{alg:alg1} with $\kappa = 0$
  and the restriction that only subsets of the distributional parameters with only a single
  element are allowed in the best-subset search. 

\item \code{BW+BS}:
  Batchwise best-subset updating with BIC, i.e., Algorithm~\ref{alg:alg1} with $\kappa = 0$. 

\item \code{BW+CF}:
  Batchwise non-cyclical updating and correlation filtering with BIC. We use
  Algorithm~\ref{alg:alg1} with the restriction that only subsets of the distributional
  parameters with only a single element are allowed in the best-subset search. 

\item \code{BW+BS+CF}:
  Batchwise best-subset updating and correlation filtering with BIC.
  We use Algorithm~\ref{alg:alg1}. 
  
 \item \code{ThresDesc}:
  Best-subset updating with threshold descent for correlation threshold selection.  
  Final model is selected with BIC.
  
\item \code{ThresDesc+BW}:
  Batchwise best-subset updating with threshold descent for correlation threshold selection. Final model is selected with BIC. 
\end{itemize}  
The different stagewise methods are summarized in Table \ref{tab:summeth}.

\begin{table}[t!]
\centering
\begin{tabular}{|l|l|l|l|}
\hline
Name & Update & Correlation Filtering & Batchwise  \\
\hline
 \code{Standard} & non-cyclical & &  \\
\hline
\code{BS} & best-subset &  &  \\
\hline
\code{CF} & non-cyclical & \checkmark &  \\
\hline
\code{BS + CF} & best-subset & \checkmark &  \\
\hline
\code{BW} & non-cyclical &  & \checkmark \\
\hline
\code{BW + BS} & best-subset &  & \checkmark \\
\hline
\code{BW + CF} & non-cyclical & \checkmark & \checkmark \\
\hline
\code{BW + BS + CF} & best-subset & \checkmark & \checkmark \\
\hline
\multirow{2}{*}{\code{ThresDesc}} & \multirow{2}{*}{best-subset} & \checkmark * with descending&  \\
&  & correlation threshold&  \\
\hline
\multirow{2}{*}{\code{ThresDesc + BW}} & \multirow{2}{*}{best-subset} & \checkmark * with descending&  \multirow{2}{*}{\checkmark}\\
&  & correlation threshold&  \\
\hline
\end{tabular}
\caption{\label{tab:summeth} Different methods used in the simulation.}
\end{table}

\subsection{Simulation Design}

We simulate data from the normal distribution (\code{NO}), the Gamma distribution (\code{GA}) and the zero-adjusted negative binomial type 1 distribution (\code{ZANBI}). All distributions used in this paper are implemented in the
\proglang{R} package \pkg{gamlss.dist} \citep{sdr:gamlss.dist}. In the simulation study,
we let all distributional parameters depend on covariates which are drawn from a uniform distribution
$x_1, \dots, x_6 \sim \mathcal{U}(-1,1)$. The complete specifications for each distribution are shown
in Table~\ref{tab:dists}.
\begin{table}[t!]
\centering
\begin{tabular}{|l|l|}
\hline
Distribution & Parameters \\
\hline
\multirow{2}{*}{$\texttt{NO}(\mu,\sigma)$} & $\mu = x_1 + 2x_2 + 0.5  x_3 - x_4$ \\
& $\log\left( \sigma \right) = 0.5x_3 + 0.25x_4 - 0.25x_5 - 0.5x_6$ \\
\hline
\multirow{2}{*}{$\texttt{GA}(\mu,\sigma)$} & $\log\left( \mu \right) = x_1 + 2x_3 + 0.5x_5 - x_6$ \\
& $\log\left( \sigma \right) = 0.5x_3 + 0.75x_4 - 0.3x_5 - 0.5x_6$ \\
\hline
\multirow{3}{*}{$\texttt{ZANBI}(\mu,\sigma,\nu)$} & $\log\left( \mu \right) = 0.5 + 0.5x_1 - x_3 + 0.75x_5 + 0.75x_6$ \\
& $\log\left( \sigma \right) = -1 + x_2 - 1.25x_4 + x_5$ \\
& $\log\left( \frac{\nu}{1 - \nu} \right) = -0.5 + x_3 - x_4 - x_5$ \\
\hline
\end{tabular}
\caption{\label{tab:dists} Distributions and specifications for the linear predictors
  of the respective parameters used in the simulation study.}
\end{table}
Here, the mean of the \code{GA} is equal to $\mu$. The variance of the \code{GA} distribution is given by $\mu^2\sigma^2$. For the \code{ZANBI}, $\mu$ is the mean of the \code{NBI} part, $\mu + \sigma\mu^2$ is the variance of the \code{NBI} part and $\nu$ corresponds to the probability of the zero event, i.e.,
$\mathbb{P}\left(Y = 0 \vert \mu,\sigma,\nu \right) = \nu$

\paragraph{Further settings}
\begin{itemize}
\item To assess the performance of our methods across both small and large data settings,
  we simulate data for varying numbers of observations including 500, 1000, 5000, 10000, 100000
  and 1000000. For the largest sample sizes of 100000 and 1000000 we limit the evaluation to
  the true model and batchwise methods only, as the computational burden of the other
  methods was deemed to be too high.
\item In addition, the influence of an additional number of noise variables (denoted with
  \code{nnoise} in the following) is considered with 30 and 100 noise variables. Each predictor
  is modeled including all available covariates. Accordingly, for each predictor
  2+\code{nnoise} or 3+\code{nnoise} non-relevant covariates are included. 
\item Two cases of correlation between the covariates are considered. Firstly, covariates
  are uncorrelated ($\rho=0$), and secondly, random pairs of covariates are correlated with
  $\rho = 0.7$. The second case is generated as follows: First, we set up a data-matrix
  with uniformly distributed entries with \code{nnobs} rows and $l$ = 6+\code{nnoise} columns.
  Second, we multiply the data-matrix with the cholesky factor $\mathbf{L}^\top$ from the
  cholesky decomposition $\mathbf{L}\mathbf{L}^\top = \mathbf{\Sigma}$ of the covariance matrix
  $$
    \mathbf{\Sigma} = \begin{pmatrix}
      1 & \rho & \rho^2 & \cdots & \rho^{l-1} \\
      \rho & 1 & \rho & \cdots & \rho^{l - 2} \\
      \vdots & \vdots & \vdots & \ddots & \vdots \\
      \rho^{l - 1} & \rho^{l - 2} & \rho^{l - 3} & \cdots & 1
    \end{pmatrix}
  $$
  to introduce the correlation to the data. At this point the correlation between neighboring
  columns is 0.7 and the correlation between columns $i$ and $j$, where
  $i,j = 1, \dots, \text{6+\code{nnoise}}$, is $0.7^{\lvert i-j\rvert}$ and thus, the correlation
  propagates from column to column. Third, we apply a random permutation to the columns. The columns
  of the final matrix correspond to the covariates $x_1,\dots,x_{6 + \text{\code{nnoise}}}$.
  The final permutation is a part of the random data generation process and therefore varies
  across different replications of the simulation. For instance, two variables may exhibit high
  correlation in one replication, while showing almost zero correlation in another replication.
\item The used threshold values ($\kappa$) depend on the number of observations or the batchsize used and is derived from the significance level $\alpha = 0.05$. Small sample sizes bear the risk of a small power for the correlation tests, i.e., not rejecting $H_0$ although $H_0$ is wrong. To circumvent this we enforce a maximum threshold $\kappa = 0.175$. On the other extreme, when the sample size is high, everything tends to be significant, thus we also enforce a minimum threshold $\kappa=0.075$. 
  
\item As mentioned before, to evaluate the performance of the methods, the (C)RPS, and the
  predictor bias, i.e., the RMSE of the estimated additive predictors and the true
  additive predictors $\text{RMSE}_k = \sqrt{\frac{1}{n} \sum_i (\hat{\eta}_{ki} -\eta_{ki})^2}$,
  for $k = \mu,\sigma, \nu$, are calculated based on an \textit{out-of-sample} validation
  data-set with $10000$ observations. Furthermore, the number of correctly selected covariates
  and the number of falsely selected covariates in each predictor (true and false positives, \code{TP} and \code{FP}) are
  counted. In addition, the computational time is tracked for all methods.
\item Each combination of the simulation design is replicated 100 times.
\end{itemize}

\begin{figure}[t!]
\centering
\includegraphics[width=1\textwidth]{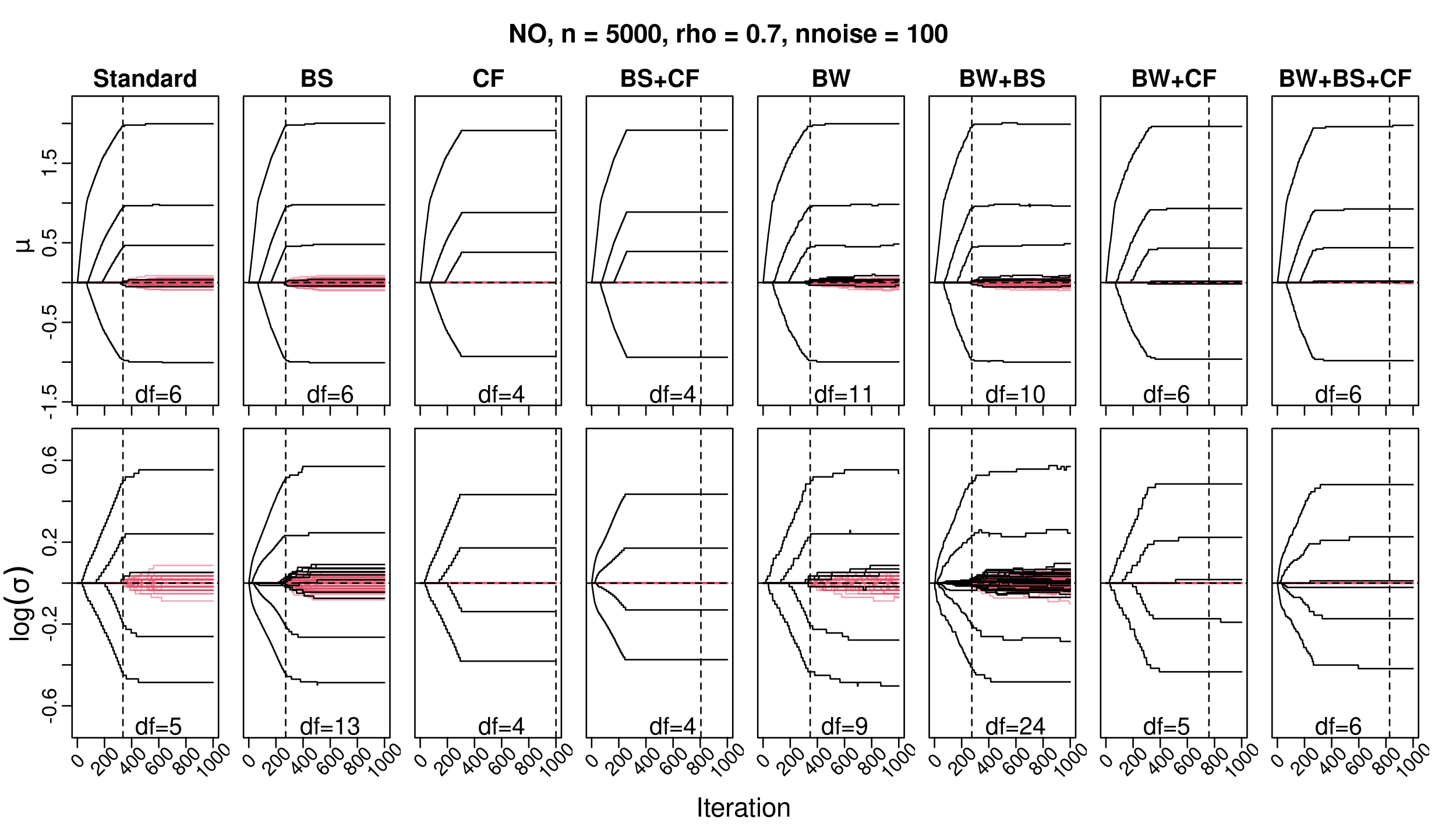}
\caption{\label{fig:comp_NO} Evolution of coefficient paths of variable selection step of new methods illustrated with the setting \code{NO}, $n=5000$, $\rho=0.7$, $\texttt{nnoise}=100$. Dashed vertical lines depict early stopping through information criteria. df is the number of selected variables for the respective distributional parameter.}
\end{figure}

Figure \ref{fig:comp_NO} illustrates the new methods, except the threshold descent method, presented using one specific simulation setting for the \code{NO}
distribution.
It can be seen that all
variants using correlation filtering have remarkably low false positive rates. However, this approach can
lead to excessive shrinkage of the coefficients, as can be observed, for example, in the method
\code{CF} coefficient-paths for parameter $\sigma$. Therefore, as mentioned earlier, the model is
re-estimated in a second step without correlation filtering and the optimal value of \code{mstop} is 
selected using the BIC.

\subsection{Simulation Results}

\begin{figure}[t!]
\centering
\includegraphics[width=0.82\textwidth]{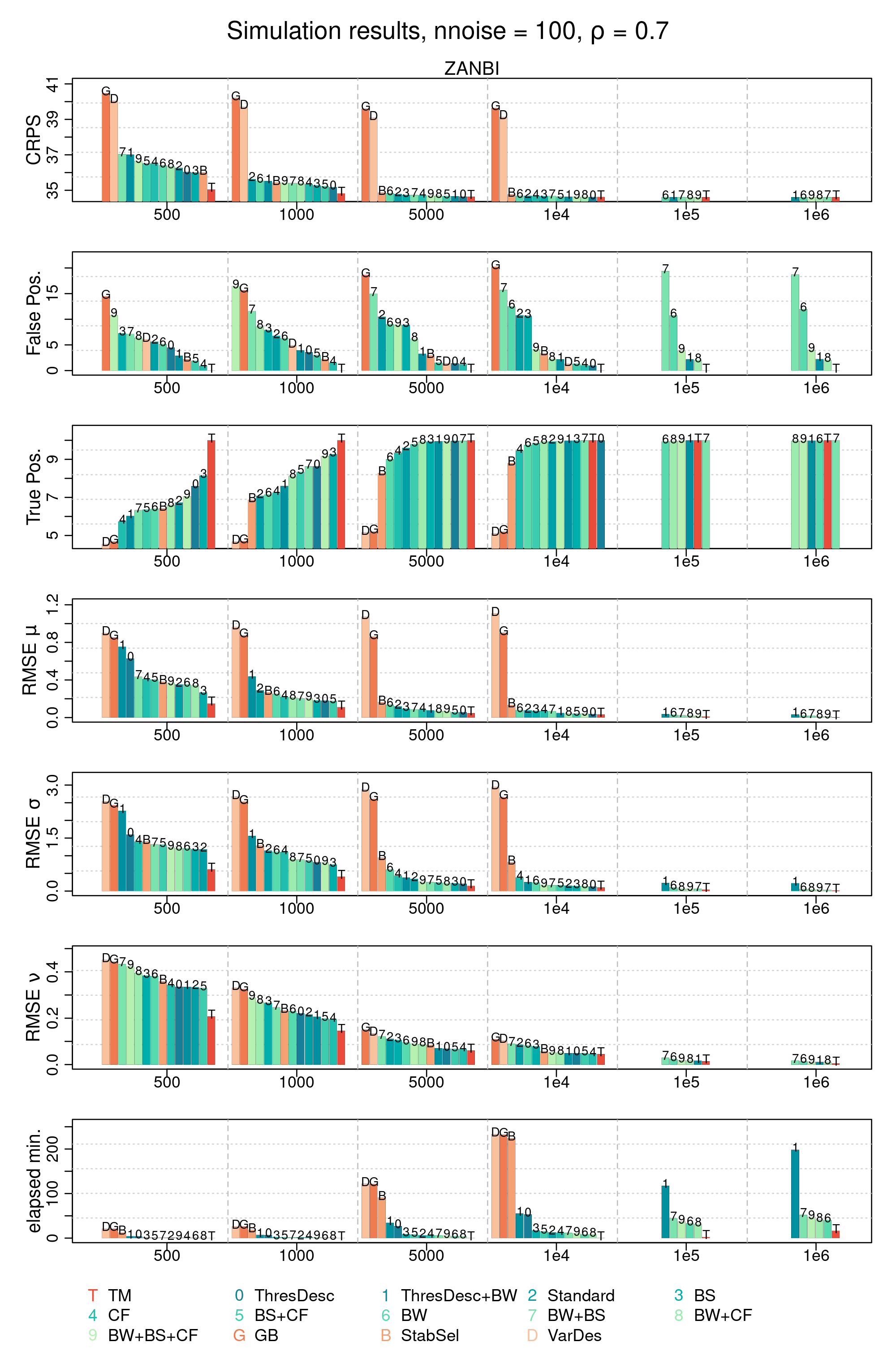}
\caption{\label{fig:res100ZANBI} Summary of all results of the simulation study for all \code{ZANBI} settings
  with $\texttt{nnoise}=100$ and $\rho = 0.7$. The average results of 100 replications
  vs.\ number of observations denoted on the abscissa is shown for the respective metrics.}
\end{figure}

We focus on presenting the simulation results of the settings of the \code{NO} and \code{ZANBI} with $\texttt{nnoise}=100$ and $\rho=0.7$ in this section. All other settings can be found in Appendix Section~\ref{app:simresults}. 
Figure~\ref{fig:res100NO} and Figure~\ref{fig:res100ZANBI} show the average results of 100 replications and we highlight these in the following:

\subsubsection{False Positives and True Positives}
For both \code{NO} and \code{ZANBI}, the comparison of false positive rates across various methods indicates that correlation filtering and \code{ThresDesc} variants consistently outperform the models \code{Standard}, \code{BS}, \code{BW}, \code{BW+BS}, \code{GB}, and \code{SAASL} in terms of false positive rate. 
In smaller settings ($n \leq 1000$), the batchwise variants (\code{BW+CF} and \code{BW+BS+CF}) tend to perform slightly worse than the full batch variants (\code{CF} and \code{BS+CF}), though this difference diminishes for settings with $n > 1000$.
Regarding true positives, \code{CF} and \code{BS+CF} perform slightly worse in selecting all true variables in the normal distribution setting with $n = 500$ compared to other methods. However, this difference disappears with larger sample sizes. In the \code{ZANBI} setting, best-subset methods combined with correlation filtering (\code{BS+CF} and \code{BW+BS+CF}) achieve a marginally better true positive rate than their non-cyclical counterparts (\code{CF} and \code{BW+CF}), which in turn achive better true positive rates than  \code{GB} and \code{SAASL} in the \code{ZANBI} settings. The latter two methods face significant issues in the challenging \code{ZANBI} scenario, exhibiting very low true positive rates and consequently high predictor bias and (C)RPS values.
The results are very promising, showing a strong ability for the correlation filtering and \code{ThresDesc} variants to perform variable selection.

\subsubsection{(C)RPS and Predictor Bias}
Regarding (C)RPS in the \code{NO} settings, all models converge to the true model (\code{TM}) as the number of observations increases. However, due to the high false positive rate, \code{GB} and \code{SAASL} models converge slightly slower than other methods. The \code{CF}, \code{BS+CF}, \code{StabSel}, and \code{ThresDesc} variants all exhibit low bias and a good (C)RPS in the \code{NO} settings.
In the \code{ZANBI} settings, \code{VarDes}, \code{GB}, and \code{StabSel} have difficulty accurately estimating some linear predictors due to the vanishing gradient problem for both small and large $n$, resulting in high (C)RPS values for \code{VarDes} and \code{GB}. On the other hand our proposed methods, which all use the semi-constant step length, are able to converge to the true linear predictors and the (C)RPS with increasing sample size.

\subsubsection{Computation Time}
The elapsed time to compute a single model is high for the methods \code{GB}, \code{VarDes} and \code{StabSel}, as the respective models have to be computed multiple times on different folds. This makes the use of these models in high data settings infeasable. This is opposite to our methods which are comparable fast on moderate sample sizes and our batchwise methods which can be computed very efficiently even in high data settings. For example the most demanding setting, \code{ZANBI} with $n = 1000000$, takes around 145 minutes for the \code{ThresDesc+BW} to compute. The \code{GB} with only $n = 10000$ already takes around 150 minutes despite the 100 times smaller dataset.  


\subsubsection{Summary}
Correlation filtering and threshold descent are highly competitive tools for variable selection, performing on par with the strong benchmark competitor \code{StabSel} in the \code{NO} settings. In the highly complex \code{ZANBI} settings, all competitors encounter some difficulties, but our proposed correlation filtering and threshold descent methods, which use semi-constant step lengths, are able to accurately model all linear predictors. Furthermore our methods do not require any costly cross-validation or other subsampling approaches, which make them computational very appealing. The batchwise variants enhance the methods even further, making them highly scalable and fast for large datasets. This makes it possible to handle datasets with millions of observations,  where all competitors quickly reach their computational limits.

\begin{figure}[t!]
\centering
\includegraphics[width=0.82\textwidth]{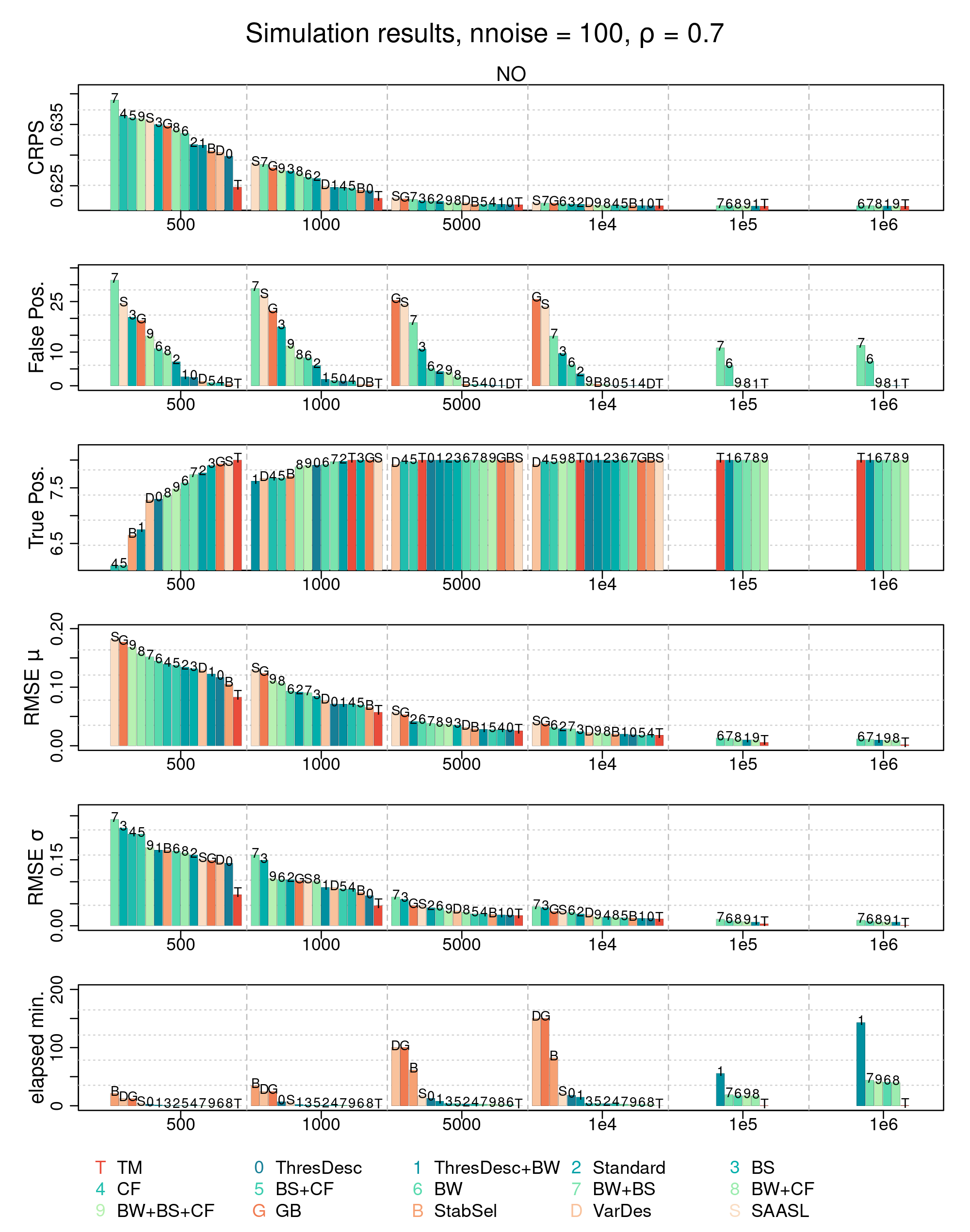}
\caption{\label{fig:res100NO} Summary of all results of the simulation study for all \code{NO} settings
  with $\texttt{nnoise}=100$ and $\rho = 0.7$. The average results of 100 replications
  vs.\ number of observations denoted on the abscissa is shown for the respective metrics.}
\end{figure}

\clearpage
\pagebreak

\section{Application: Lightning Forecast in Austria} \label{sec:flash}

Lightning is a natural phenomenon that arises from imbalances in atmospheric electrical charge
during thunderstorms. The resulting electrical discharge can cause significant damage to property
and pose a danger to people and animals \citep{Yair_2018}. In addition to its immediate effects,
lightning also has important implications for the global climate, as it is a major source of
atmospheric nitrogen oxides \citep{schumann2007}, a potent greenhouse gas that can affect
climate change \citep[Figure SPM.2 in][]{ipcc2021}. However, the relationship between lightning
and climate change remains a subject of scientific debate \citep{murray2018}, in part
because lightning processes are not yet fully understood and cannot be resolved by weather forecast models used to predict the state of the atmosphere. As a result, many studies of lightning rely on proxies such as
cloud top height \citep{price1992}, ice flux in the mid atmosphere \citep{finney2014},
or wind shear \citep{taszarek2021}, that are based on fairly simple formulation and consider only
certain aspects of the physical processes involved.

To address these limitations, scholars have proposed machine learning (ML) approaches that
incorporate multiple physical processes in their analysis of lightning
\citep[e.g.,][]{ukkonen2019, Simon2023}. These approaches are capable of processing
large amounts of data and identifying the most relevant variables from a pool of inputs, but
typically focus on describing the occurrence of lightning via binary classification rather than on
the number of lightning counts, which is a crucial variable for investigating flash rates
\citep{cecil2014}. By using ML to gain a more nuanced understanding of lightning processes,
researchers may be better equipped to elucidate the relationship between lightning and climate
change and to develop more accurate models for predicting lightning behavior in the future.

The stagewise boosting algorithm we propose is ideally suited for modeling flash rates. Firstly, it 
enables variable selection even with very large datasets. Additionally, its numerical stability is 
crucial, particularly when dealing with complex distributions such as the count data distribution employed in this application.

\paragraph{Data}
We analyze lightning counts using high-resolution data from the Austrian Lightning Detection
and Information System (ALDIS, \citealp{schulz2005}) and reanalysis data from ERA5, the fifth
generation of atmospheric reanalyses from the European Centre for Medium-Range Weather Forecasts
(ECMWF, \citealp{era5a, era5b}). ERA5 provides hourly globally complete and consistent
pseudo-observations of the atmosphere with a horizontal resolution of approximately
$32~\mathrm{km}\times 32~\mathrm{km}$, spanning from 1950 to the present. Our study not only provides a comprehensive
description of lightning but also supports a complete reanalysis to study climate trends in
lightning \citep[c.f.][]{Simon2023}. This is particularly important as homogeneous lightning
observations from ALDIS are available only for the period from 2010 to 2019. In total, our dataset
includes more than 9.1 million observations, which is particularly challenging for the estimation
of complex distributional regression models. We use the data corresponding to 2010 up to 2018 ($\approx 8.2$ million observations) as training data and the year 2019 as validation data.

\paragraph{Model Specification}
Considering that only approximately $2.65\%$ of our observations exhibit positive counts, we employ a zero-adjusted negative binomial model (\code{ZANBI}) distribution to model lightning counts.
We include $84$ physical variables from ERA5 in our analysis, which we initially scale using a smooth kernel density estimate of the empirical distribution function to ensure that all variables fall within the interval $[0, 1]$. 
Following this initial transformation, we augment each variable with eight different transformations,
resulting in each linear predictor having a pool of $84\cdot 8 = 672$ variables to select from:
\begin{equation*}
\begin{aligned}[c]
&x \mapsto x \\
&x \mapsto x^3 \\
&x \mapsto \exp{(x)} \\
&x \mapsto \mathrm{logm}(x) =\log{(1-x+0.01)}
\end{aligned}
\qquad
\begin{aligned}[c]
&x \mapsto x^2\\
&x \mapsto \sqrt{x}\\
&x \mapsto \mathrm{logp}(x)=\log{(x+0.01)}\\
&x \mapsto \mathrm{logitc}(x)=\log{\left(\frac{0.999\cdot x + 0.001\cdot 0.5}{1-(0.999\cdot x + 0.001\cdot 0.5)} \right)}
\end{aligned}
\end{equation*}

The resulting variables are then standardized as the last variable preparation step.
To better capture the characteristics of rare positive lightning events, we aim to improve our model
by subsampling the zero count data during the variable selection and refitting steps. We achieve this by 
specifying batches consisting of $10000$ random samples each from the lightning $\texttt{counts}=0$
and $\texttt{counts}>0$ observations. 
This gives us a batch with 20000 observations for every iteration step. By carrying out the batchwise correlation filtering with these subsampled batches, we can adjust for the rarity of positive lightning
events in our dataset. In contrast, the approach taken by
\citealp{sdr:Simon+Mayr+Umlauf+Zeileis:2019} is considerably more complex. They use a two-stage
hurdle model, consisting of a binomial model for lightning occurrence (yes/no) and a zero-truncated
negative binomial model for positive counts. They also use stability selection with boosting to
select the most important variables for their model, which is computationally very demanding.

To account for the subsampling, we apply an intercept adjustment in the logistic regression part
of the model afterwards. This correction, derived from \citealp{King2001}, ensures that the corrected
subsampled distribution is consistent with the original distribution. In the Appendix
Section~\ref{app:intadj} we provide a proof that the parameter estimates do not require
modification, except for the logistic regression intercept of the \code{ZANBI} model $\boldsymbol{\beta}_{0\nu}$. 
The subsampled parameter
$\boldsymbol{\beta}_{0\nu}^{\text{sub}}$ gets adjusted to
$$
\boldsymbol{\beta}_{0\nu} = \boldsymbol{\beta}_{0\nu}^{\text{sub}} -
  \log\left( \frac{1-\tau_0}{\tau_0}\cdot \frac{t_0}{1-t_0}\right),
$$
where $\tau_0$ is the proportion of zeros in the full dataset and $t_0$ is the proportion of zeros in
the subsampled dataset (i.e., in each batch). We have $t_0 = 0.5$ and $\tau_0 \approx 0.0265$.

We include all 672 variables as predictors and we follow a two-step
procedure. The first step involves variable selection using correlation filtering and best-subset
selection. The second step involves refitting with best-subset selection until convergence
using the variables selected in step one.

\paragraph{Results}

\begin{figure}[t!]\centering
\includegraphics[width=0.95\textwidth]{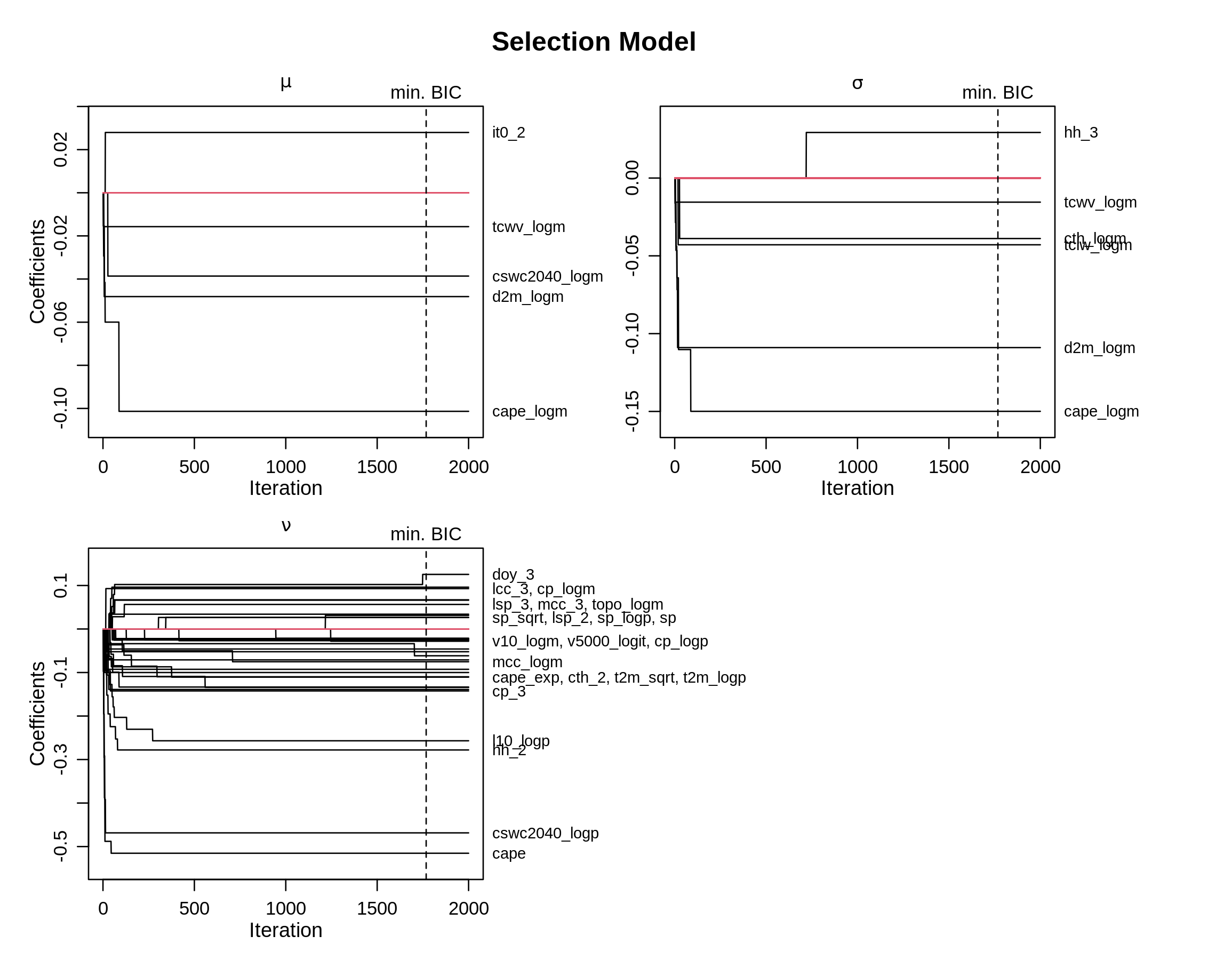}
\caption{\label{fig:coefs_sel} \code{ZANBI} Variable selection process. Shown are the coefficient paths of the selected variables.} 
\end{figure}

Once variables are selected, the best-subset
stagewise boosting algorithm is applied to refit the model until convergence. The selection process is depicted in Figure~\ref{fig:coefs_sel}. It shows a balanced selection of variables for every linear predictor.
The variable importance score of the selected variables for four years is presented in Figure~\ref{fig:vis_4y} and for all years in the appendix Section~\ref{app:vis}. The algorithm selected various transformations of 22 variables into the final model, with \code{cswc2040}, the mass of specific snow water content between the $-20^\circ$C and $-40^\circ$C isotherms, being the most important one according to the variable importance score (explained in caption of Figure~\ref{fig:vis_4y}). Other important variables are \code{mcc}, the medium cloud cover, \code{cp}, the convective potential and \code{cape}, the convective available potential energy.
Note that the variable importance appears to differ for a number of variables in the validation data, such as \code{cape}, in comparison to the scores observed in the data used for model estimation. This also illustrates the difficulty of developing a forecasting model for lightning counts, given the complex physical phenomena that drive the data.

\begin{figure}[t!]\centering
\includegraphics[width=0.95\textwidth]{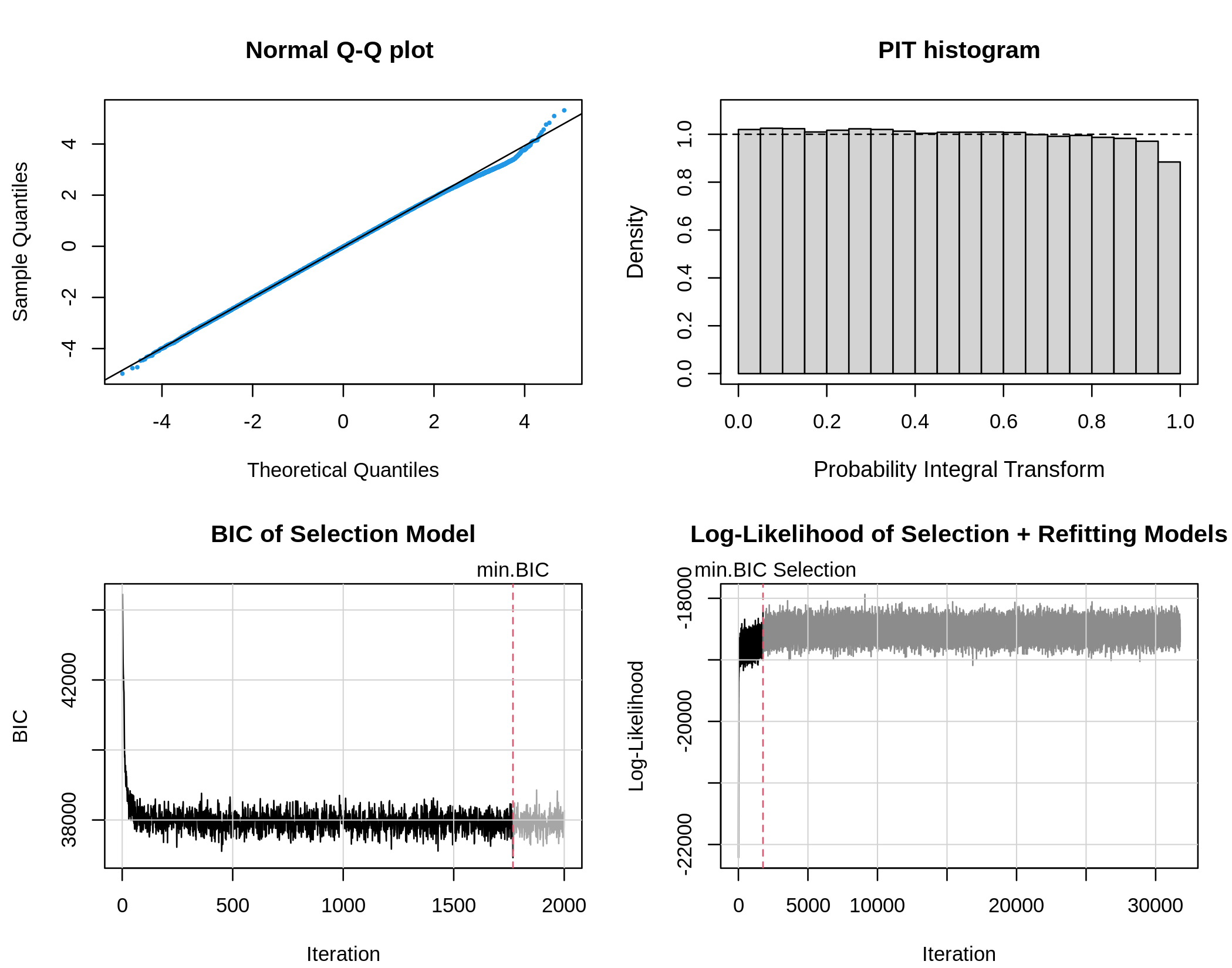}
\caption{\label{fig:diag} Diagnostic plots are shown for the \code{ZANBI} lightning model. The top row shows a quantil-quantil plot (top left) and a probability integral transform plot (top right) for out-of-sample observations from the year 2019. The BIC of the selection model is shown on the bottom left. The bottom right shows the in-sample log-likelihood path of the selection model and the refitting model,  where the vertical red lines denote the early stopping in the selection step induced by BIC. } 
\end{figure}

Results in Figure~\ref{fig:diag} based on the out-of-sample data show firstly, a QQ-plot based on randomized quantile residuals \citep{sdr:Dunn+Smyth:1996}, indicating a quite good fit for all but some extreme values corresponding to observations with very high counts. Secondly, the PIT histogram also indicates a overall well calibrated model fit. The bottom row of Figure~\ref{fig:diag} shows the BIC curve of the selection model with the minimum at iteration 1768 and a log-likelihood plot of the selection model and the refitting model. At the transition from the selection to the refitting model, the log-likelihood plot experience a rapid jump, before it reaches a stable level for almost 30000 iterations. 

\begin{figure}[t!]
\centering
\includegraphics[width=1\textwidth]{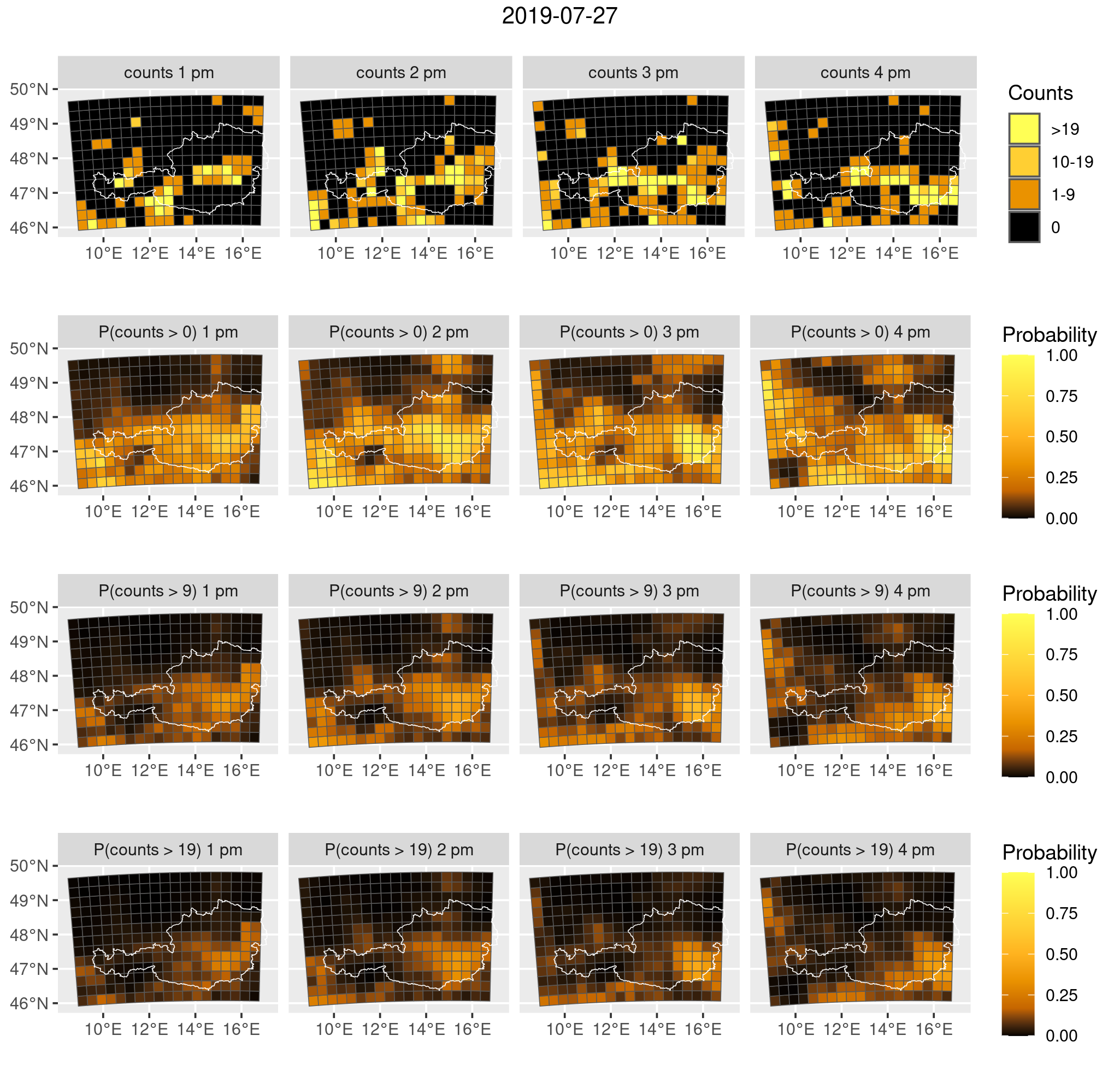}
\caption{\label{fig:Spatial} \code{ZANBI} lightning model. Compared is a (out of sample) forecast for the date 2019-07-27 with the observed counts. The time evolution from 1 pm up to 4 pm is depicted along the horizontal axis in the panel plot. The top row shows the observed number of lightning counts, the second row shows the probability for at least one count, the third row shows the probability for at least 10 counts and the fourth row shows the probability for at least 20 counts. }
\end{figure}

Figure~\ref{fig:Spatial} exemplary shows a forecast based on the estimated model for Austria on the afternoon (1 pm to 4 pm, local time) on July 27, 2019. Whilst the first row shows the observed number of lightning strikes, rows 2-4 show the probability of exceeding 0,9 and 19 strikes. It can be seen that the model is able to properly depict areas where high lightning counts occur.

\begin{figure}[t!]\centering
\includegraphics[width=0.95\textwidth]{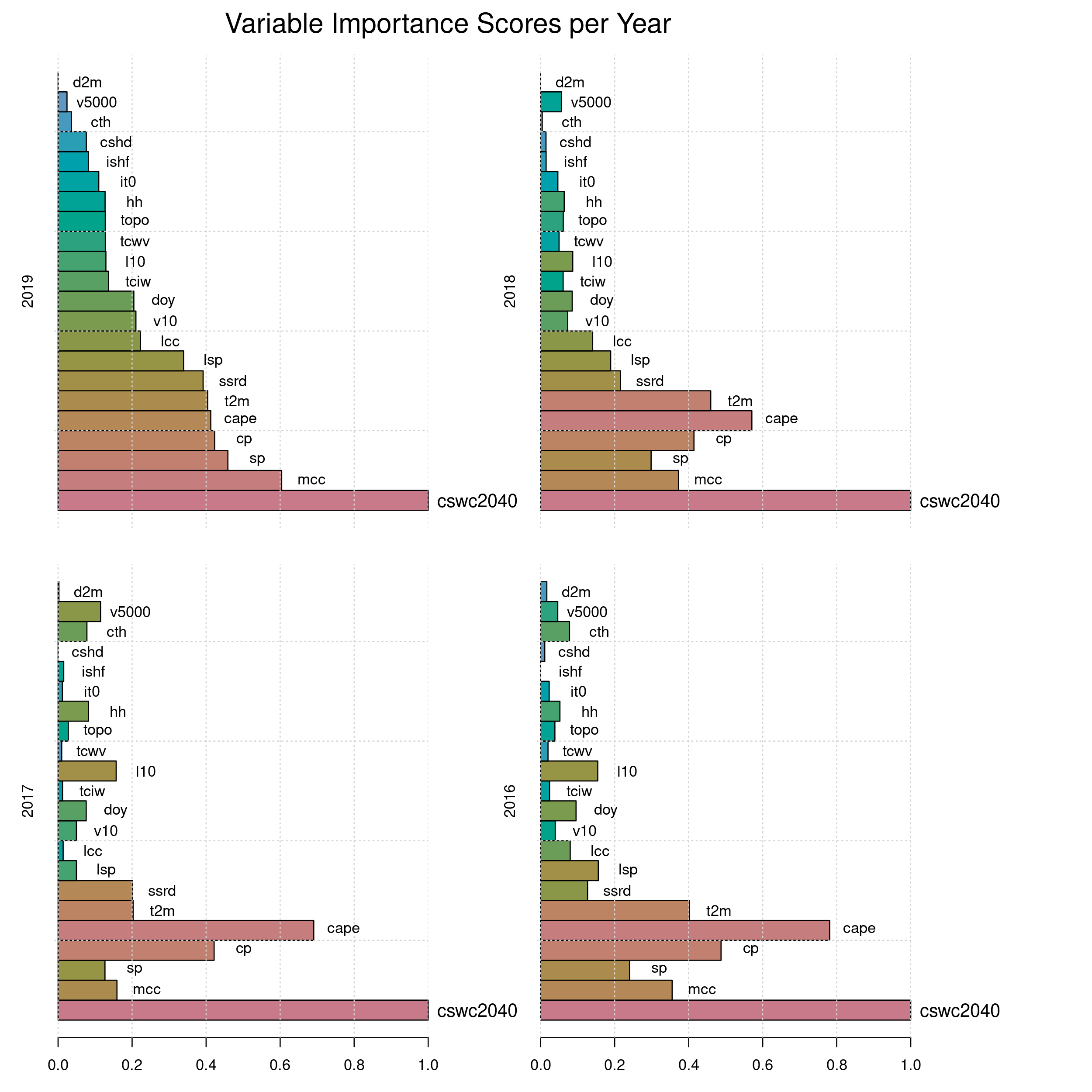}
\caption{\label{fig:vis_4y} \code{ZANBI} lightning model. Variable importance scores (VIS) - VIS are calculated using a permutation-based approach (similar to the permutation-based VIM of \citet{WEI2015}). For each variable, the log-likelihood of the model is compared to the log-likelihood of the model where the values of that variable in the year-specific subset of the data are permuted. The resulting scores represent the change in log-likelihood attributable to each variable. Scores are then normalized to the range [0, 1], with 0 indicating minimum impact and 1 indicating maximum impact on the model's predictive performance for the respective year. The model is trained using data from 2010 to 2018, with data from 2019 serving as the true validation dataset.} 
\end{figure}

\paragraph{Summary}
Our algorithm successfully identifies relevant variables for the prediction of lightning events in Austria. By using the stagewise 
boosting approach, a complex distributional regression model could be estimated with this large data set of over 8.2 million observations. 
The entire estimation process takes about 11 1/2 hours (selection step: 4h 25min; refitting step: 7h 12min). This processing 
time is considerably faster compared to other methods, for example if stability selection would be used, and underlines the efficiency 
of our algorithm. Furthermore, the numerical stability of the algorithm is crucial for dealing with the zero-adjusted negative binomial 
model for flash counts, as it ensures accurate and reliable variable selection.

\clearpage
\pagebreak

\section{Summary}

Our contributions include the concept of stagewise boosting distributional regression, which offers 
clear advantages over standard gradient boosting for the estimation of complex models. This is 
achieved by using a semi-constant step length to avoid the vanishing gradient problem. Furthermore, 
we introduce a novel updating scheme, best subset updating, which iteratively updates the best 
subsets of the possible distribution parameters. Additionally, we introduce a new variable selection method, 
correlation filtering, which proves to be very effective in high-dimensional data settings without 
the need for costly cross-validation or other subsampling approaches, making it computationally very 
attractive. Our proposed algorithm can be further optimized applying a batchwise variant for big 
data problems. This variant combines the principles of stochastic gradient descent and component-
wise boosting to compute update steps on randomly selected subsets of the data. This approach 
enables the estimation of models with very large data sets and reduces the risk of getting stuck in 
local optima.

In an extensive simulation study, we demonstrate that our methods, evaluated with metrics such as false positive rates and CRPS, are at least equivalent to other benchmark methods, but are much faster, because no 
complex methods for, e.g., finding the optimal stop iteration are necessary, and also no complex 
resampling. Furthermore, we show that stagewise boosting has clear advantages in more demanding 
situations, e.g., when estimating a zero-adjusted negative binomial distribution. The latter 
distribution was used in a challenging lightning forecasting application, which incorporates over 
9.1 million observations with 672 possible variables to choose from for each of the three 
distributional parameters of the \code{ZANBI} distribution.

One aspect not considered in this work is the inclusion of regression splines instead of purely linear effects. In the future, we plan to integrate these into the algorithm to enable the estimation of spatial effects, for example.

\section*{Acknowledgements}

This project was partially funded by the Austrian Science Fund~(FWF) grant number~$33941$.
We are grateful for data support by Gerhard Diendorfer and Wolfgang Schulz from OVE-ALDIS.
The computational results presented here have been achieved (in part) using the
LEO HPC infrastructure of the University of Innsbruck.

\bibliography{stagewise.bib}

\section{Appendix}

\subsection{Proof of Consistency of Subsampling Correction and Intercept Adjustment} \label{app:intadj}

\citealp{King2001} prove the consistency of the subsampling correction.  Following their notation, let us consider random variables $X, Y$ with density $\mathbb{P}(X,Y)$ representing the full sample and random variables $x,y$ with density $\mathbb{P}(x,y)$ representing the subsampled case, with all positives and only a random selection of zeros from $X,Y$. Furthermore let $d$ and $D$ be random samples of size $n$ drawn from $\mathbb{P}(X,Y)$ and $\mathbb{P}(x,y)$ respectively. In this setting, they showed, as $n \rightarrow \infty$:
$$ \mathbb{P}(y\vert x,d)\cdot \underbrace{\frac{\mathbb{P}(Y\vert D)}{\mathbb{P}(y\vert d)} \cdot \frac{\mathbb{P}(x\vert d)}{\mathbb{P}(X\vert D)} }_{\text{subsampling correction}}\stackrel{d}{\rightarrow} \mathbb{P}(Y\vert X)$$
and similar
$$ \mathbb{P}(Y\vert X,D)   \stackrel{d}{\rightarrow} \mathbb{P}(Y\vert X),$$
Thus, the corrected subsampled distribution is consistend for the distribution of interests.
For the intercept adjustment in a zero adjusted count data model the general probability mass function reads:

\begin{align*} 
\mathbb{P}{(y = j\vert x,d)} &= \begin{cases}
             \nu(y = 0\vert x,d)  & \text{if } j = 0 \\
             (1-\nu(y = 0\vert x,d))\cdot f(y = j\vert x,d)  & \text{else},
       \end{cases}
\end{align*}
where $\nu(y = 0\vert x,d) = (1+\mathrm{exp}(-x_k\beta))^{-1}$ is the logistic regression function and $f(\cdot)$ is a zero truncated count distribution.
We abbreviate the correction factor $\frac{\mathbb{P}(Y= j\vert D)}{\mathbb{P}(y = j\vert d)} = \frac{\tau_j}{t_j}$ and rewrite $\frac{\mathbb{P}(x\vert d)}{\mathbb{P}(X\vert D)} = (\sum_{j = 0}^{\infty} \mathbb{P}(y = j\vert x,d)\cdot \frac{\tau_j}{t_j})^{-1}$. Furthermore, assume we only subsample the zeros, then $\frac{\tau_j}{t_j} = \frac{1-\tau_0}{1-t_0}$ for $j >0$. The corrected probabilities are then:
\begin{align*} 
&\mathbb{P}(Y = 0\vert X,D) = \mathbb{P}(y= 0\vert x,d)\cdot \underbrace{ \frac{\tau_0}{t_0}  \cdot (\sum_{j = 0}^{\infty} \mathbb{P}(y = j\vert x,d)\cdot \frac{\tau_j}{t_j})^{-1} }_{\text{subsampling correction}}\\
=&\frac{\mathbb{P}(y= 0\vert x,d)\cdot \frac{\tau_0}{t_0} }{\mathbb{P}(y= 0\vert x,d)\cdot \frac{\tau_0}{t_0} + (1-\mathbb{P}(y= 0\vert x,d))\cdot \frac{1-\tau_0}{1-t_0}}\\
=& \left(1+\frac{t_0}{\tau_0}\cdot\frac{1-\tau_0}{1-t_0} \cdot \frac{1}{\mathbb{P}(y= 0\vert x,d)}-1 \right)^{-1}\\
=& \left(1+\mathrm{exp}\left(-x_k\beta + \log\left(\frac{t_0}{\tau_0}\cdot\frac{1-\tau_0}{1-t_0}\right)\right)\right)^{-1}
\end{align*} 
and for $i >0$:
\begin{align*} 
&\mathbb{P}(Y = i\vert X,D) = \mathbb{P}(y= i\vert x,d)\cdot \underbrace{\frac{1-\tau_0}{1-t_0}  \cdot (\sum_{j = 0}^{\infty} \mathbb{P}(y = j\vert x,d)\cdot \frac{\tau_j}{t_j})^{-1}}_{\text{subsampling correction}} \\
=&\frac{\mathbb{P}(y= i\vert x,d)\cdot \frac{1-\tau_0}{1-t_0} }{\mathbb{P}(y= 0\vert x,d)\cdot \frac{\tau_0}{t_0} + (1-\mathbb{P}(y= 0\vert x,d))\cdot \frac{1-\tau_0}{1-t_0}}\\
=& \left(\frac{1-\mathbb{P}(y = 0\vert x,d)}{\mathbb{P}(y = i\vert x,d)}+\frac{\tau_0}{1-\tau_0}\cdot \frac{1-t_0}{t_0} \cdot \frac{\mathbb{P}(y = 0\vert x,d)}{\mathbb{P}(y = i\vert x,d)} \right)^{-1}\\
=& f(y = i\vert x,d)\left(1+\frac{\tau_0}{1-\tau_0}\cdot \frac{1-t_0}{t_0} \cdot \frac{\mathbb{P}(y = 0\vert x,d)}{(1-\mathbb{P}(y = 0\vert x,d))}\right)^{-1}\\
=& \left(1-\left(1+\mathrm{exp}\left(-x_k\beta + \log\left(\frac{t_0}{\tau_0}\cdot\frac{1-\tau_0}{1-t_0}\right)\right)\right)^{-1} \right) \cdot f(y = i\vert x,d).
\end{align*}
In both cases only a constant term was added to the linear predictor of the logistic regression part and the estimates of the non intercept variables are not changed. This is in effect an intercept adjustment. 

 \FloatBarrier

\clearpage
\subsection{Flash Model - VIS all years} \label{app:vis}
\FloatBarrier
\begin{figure}[H]\centering
\includegraphics[width=0.875\textwidth]{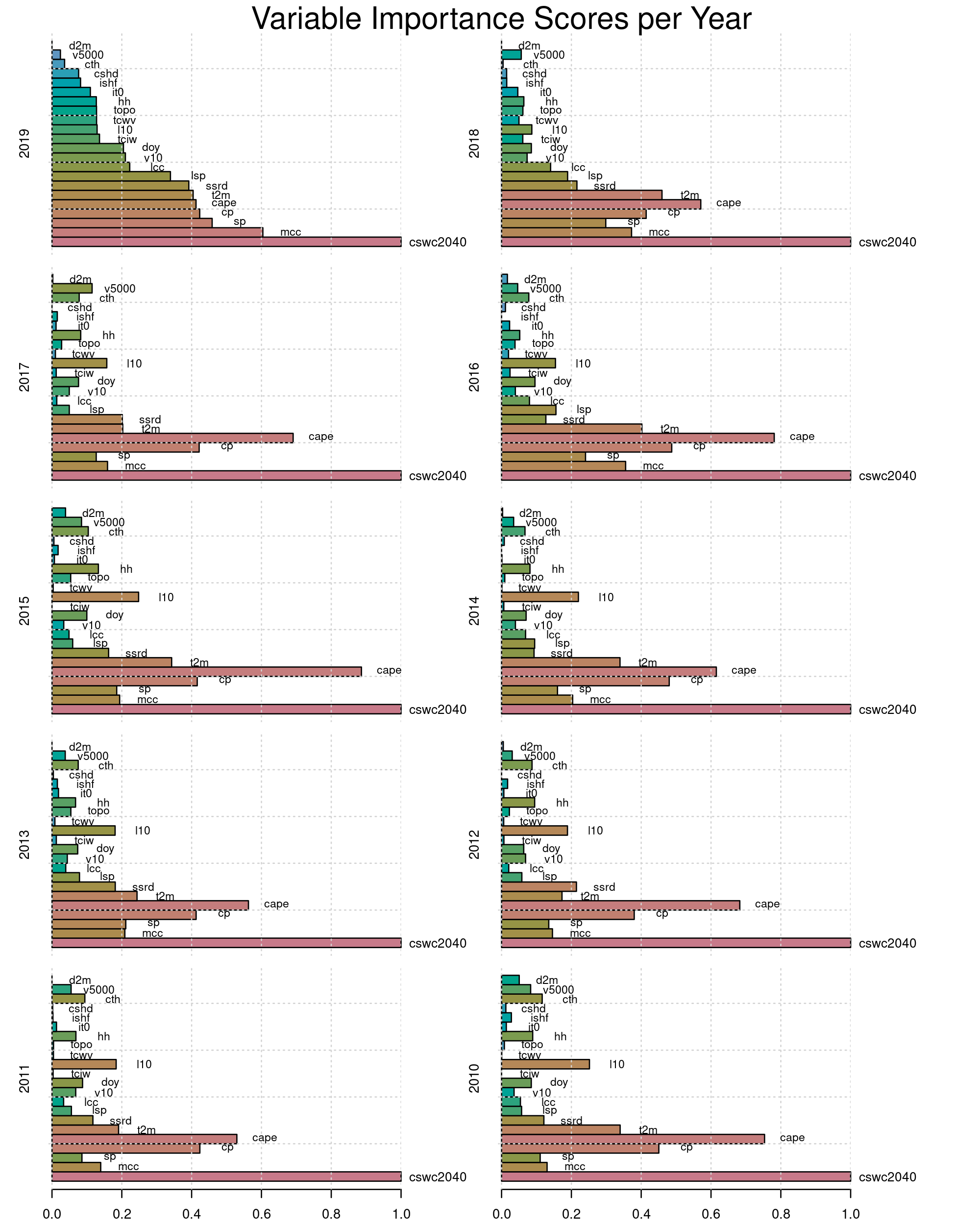}
\caption{\label{fig:vis_all} \code{ZANBI} lightning model. Variable importance scores (VIS) - VIS are calculated using a permutation-based approach. For each variable, the log-likelihood of the model is compared to the log-likelihood of the model where the values of that variable in the year-specific subset of the data were permuted. The resulting scores represent the change in log-likelihood attributable to each variable. Scores are then normalized to the range [0, 1], with 0 indicating minimum impact and 1 indicating maximum impact on the model's predictive performance for the respective year. The model was trained using data from 2010 to 2018, with data from 2019 serving as the true validation dataset.} 
\end{figure}

 \FloatBarrier

\subsection{Simulation Results} \label{app:simresults}
\FloatBarrier

\begin{figure}[H]
\centering
\includegraphics[width=0.82\textwidth]{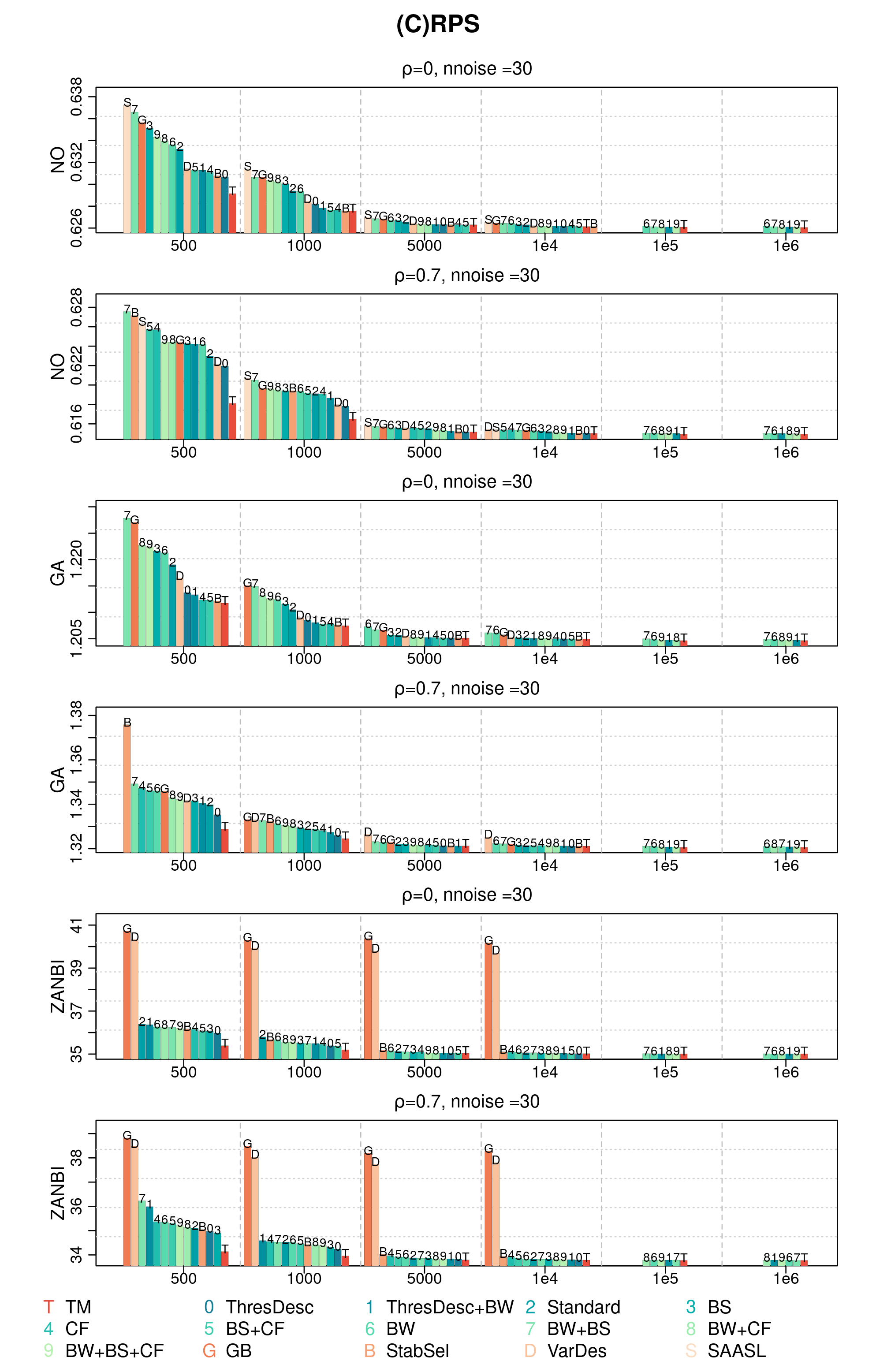}
\caption{\label{fig:crps1} Simulation study. Average (C)RPS rates of 100 replications vs. number of
observations for the different settings.}
\end{figure}

\begin{figure}[t!]
\centering
\includegraphics[width=0.82\textwidth]{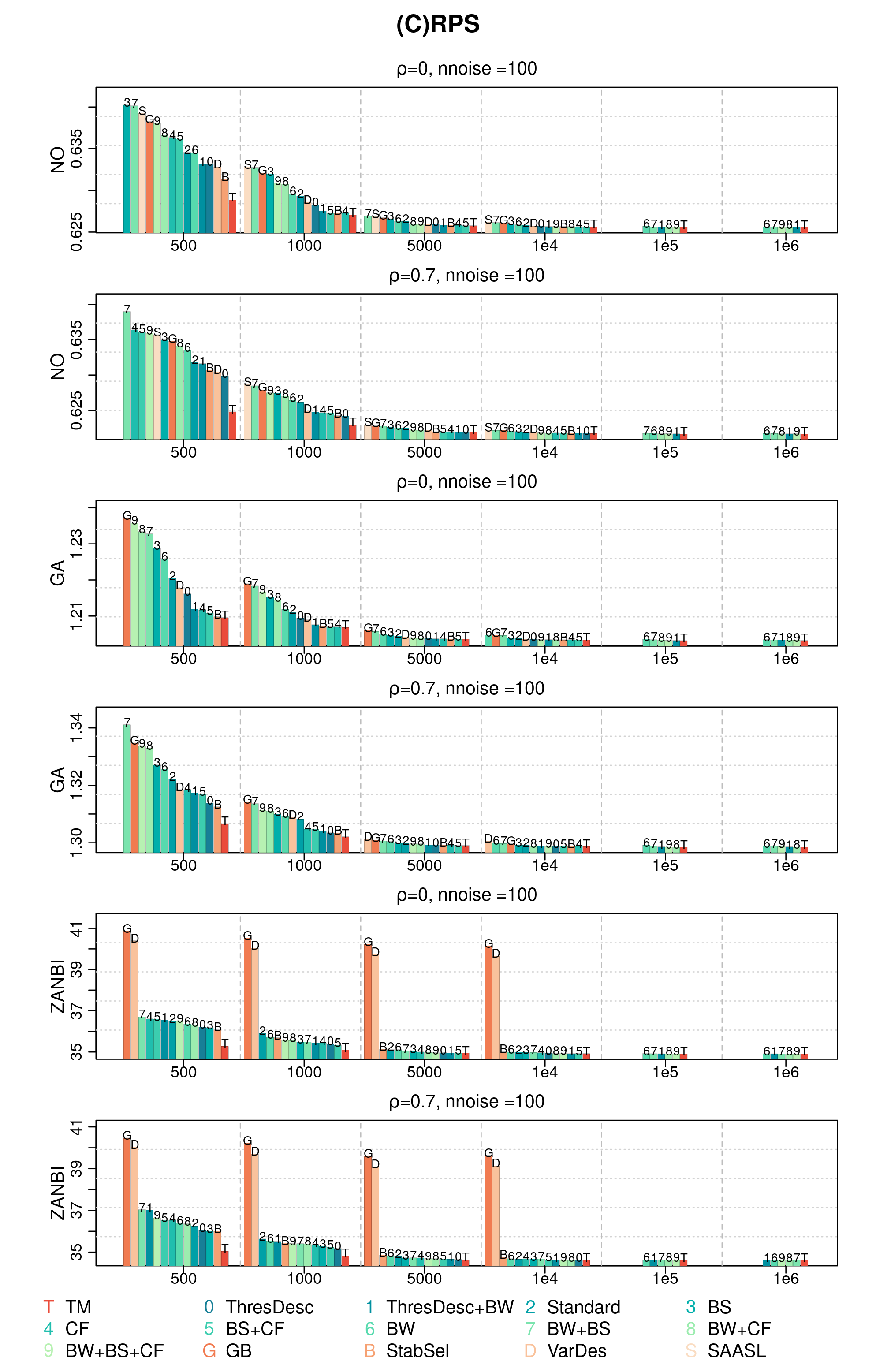}
\caption{\label{fig:crps2} Simulation study. Average (C)RPS rates of 100 replications vs. number of
observations for the different settings.}
\end{figure}

\begin{figure}[t!]
\centering
\includegraphics[width=0.82\textwidth]{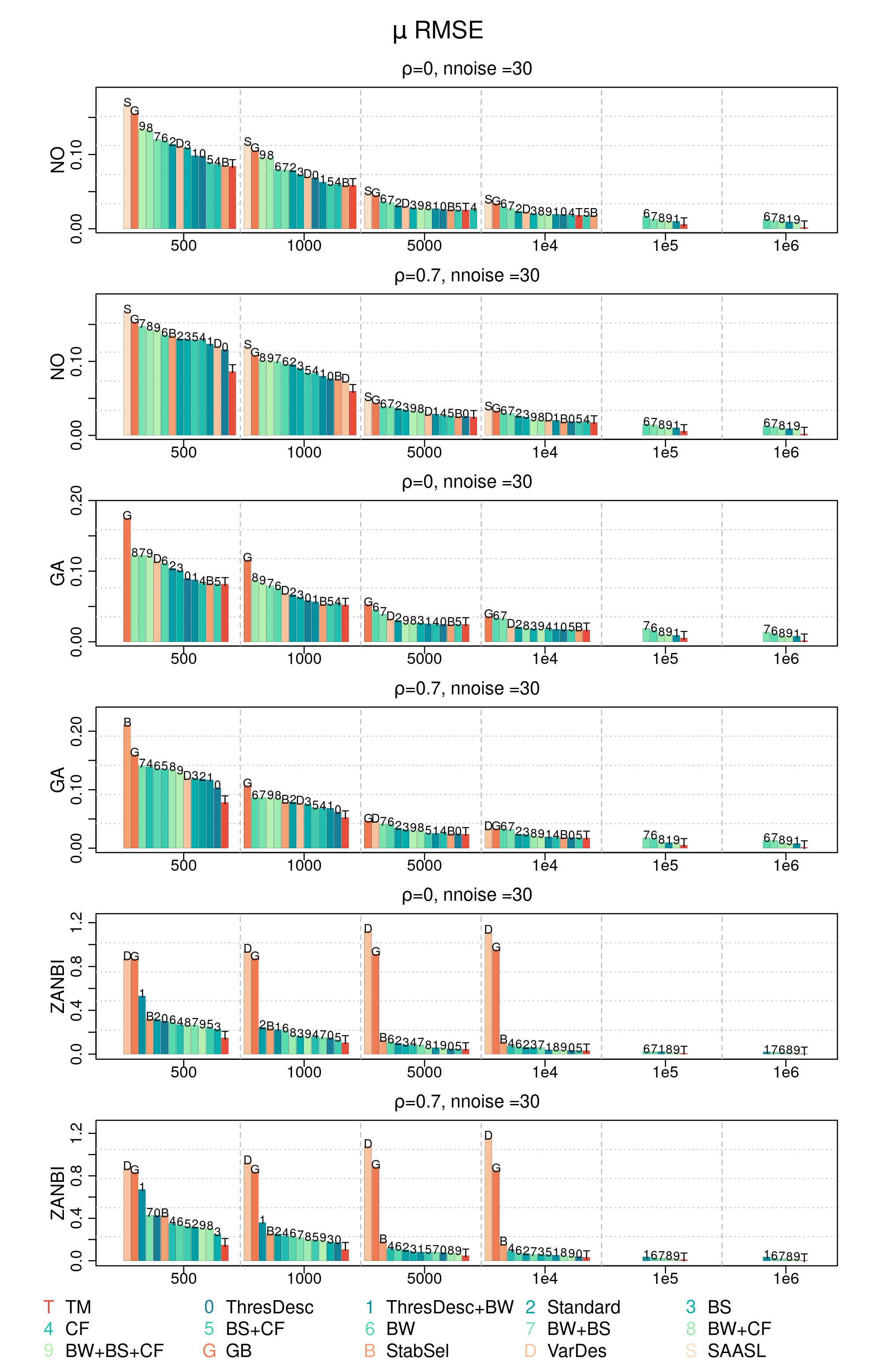}
\caption{\label{fig:mu1} Simulation study. Average RMSE of the $\mu$-predictor of 100 replications vs. number of
observations for the different settings.}
\end{figure}

\begin{figure}[t!]
\centering
\includegraphics[width=0.9\textwidth]{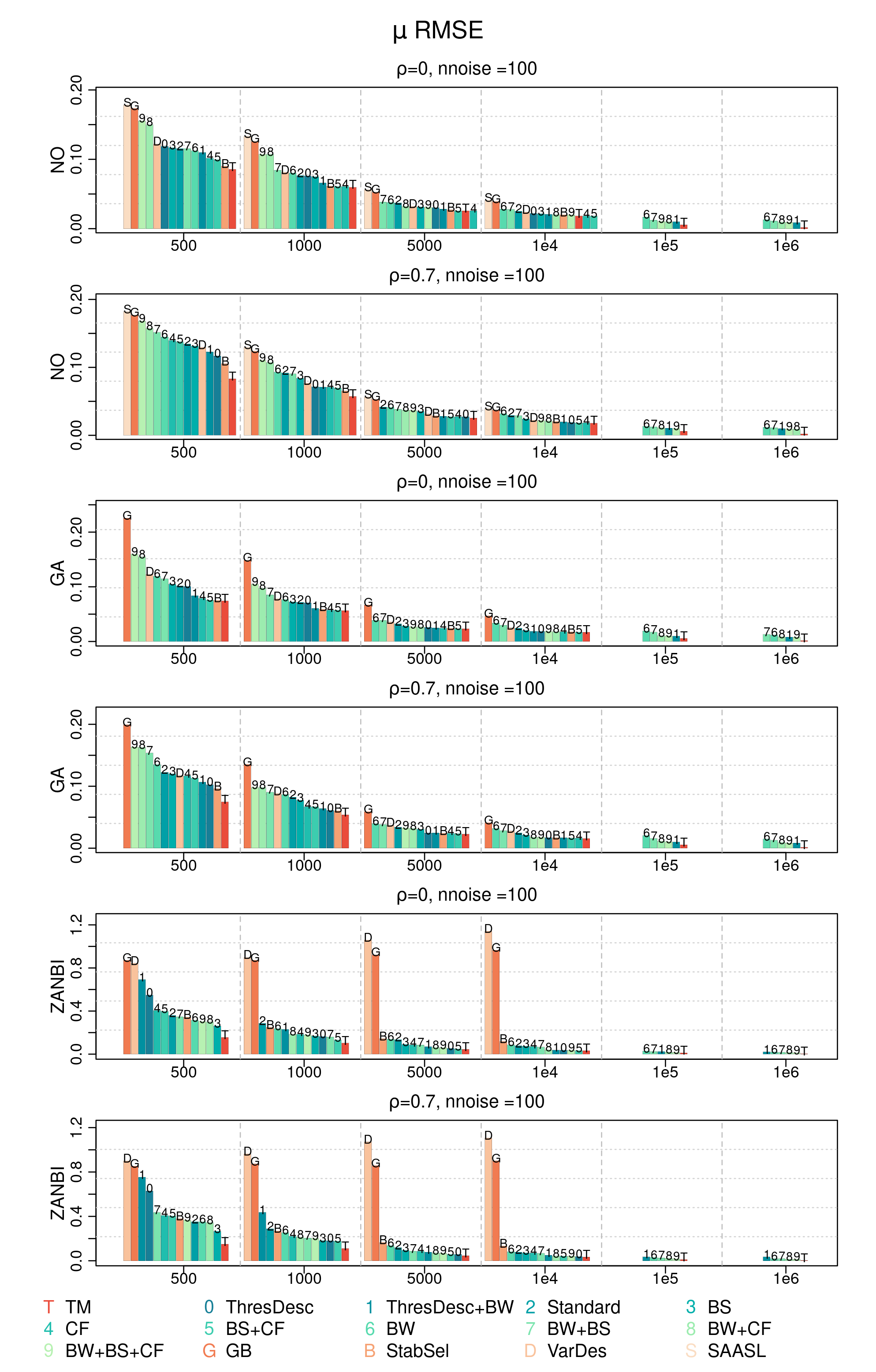}
\caption{\label{fig:mu2} Simulation study. Average RMSE of the $\mu$-predictor of 100 replications vs. number of
observations for the different settings.}
\end{figure}

\begin{figure}[t!]
\centering
\includegraphics[width=0.82\textwidth]{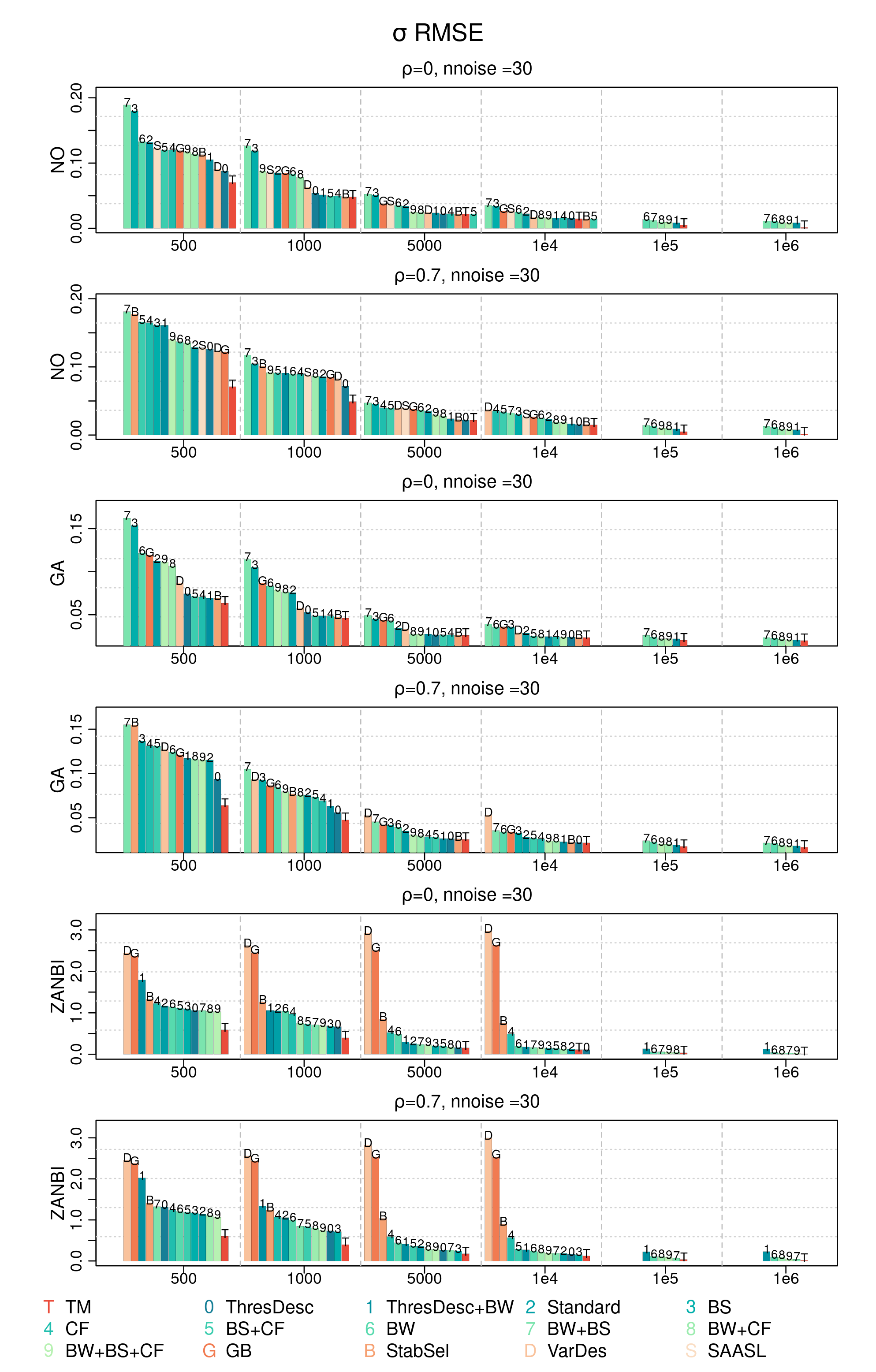}
\caption{\label{fig:sigma1} Simulation study. Average RMSE of the $\sigma$-predictor of 100 replications vs. number of
observations for the different settings.}
\end{figure}

\begin{figure}[t!]
\centering
\includegraphics[width=0.82\textwidth]{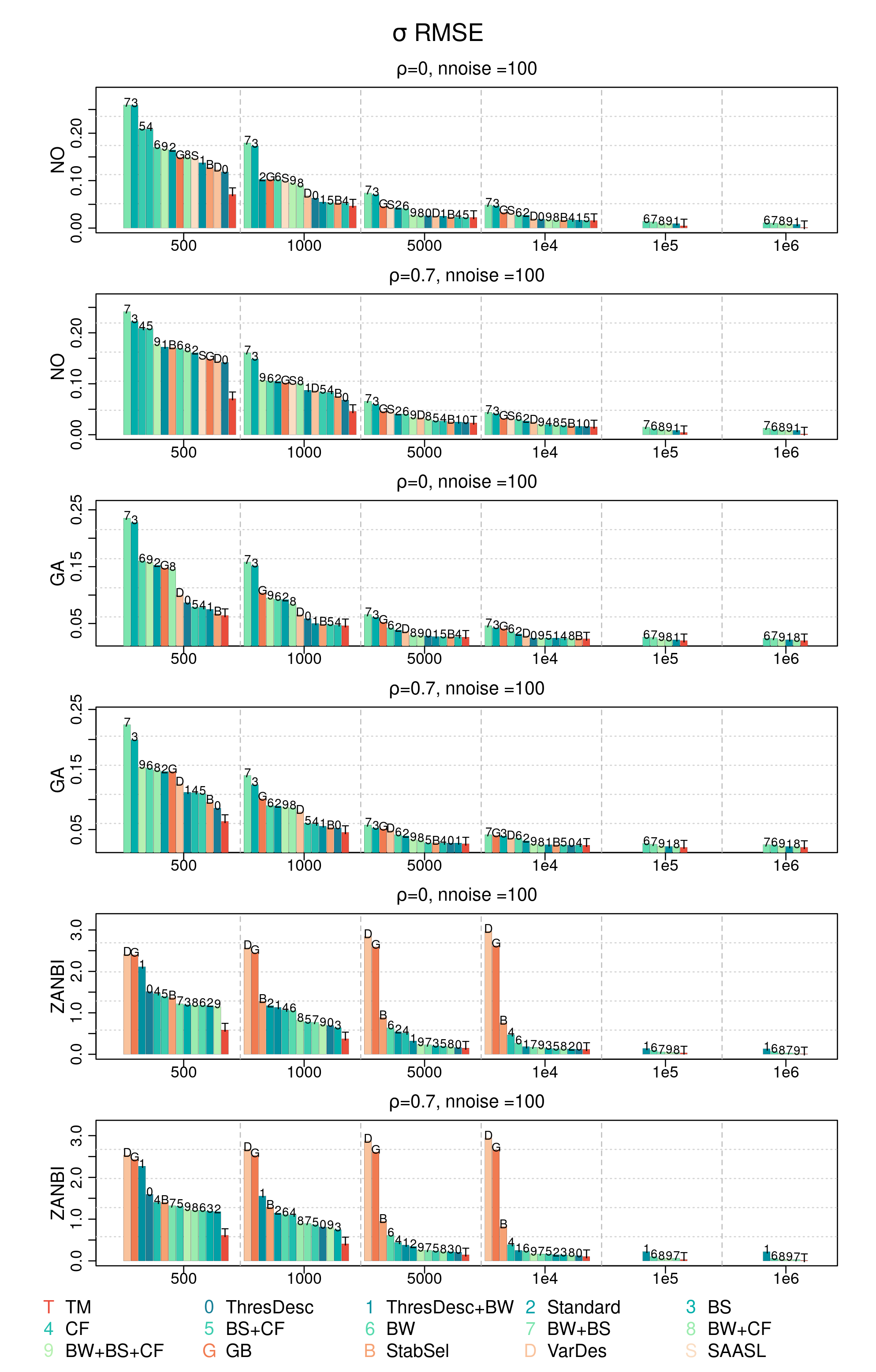}
\caption{\label{fig:sigma2} Simulation study. Average RMSE of the $\sigma$-predictor of 100 replications vs. number of
observations for the different settings.}
\end{figure}

\begin{figure}[t!]
\centering
\includegraphics[width=0.82\textwidth]{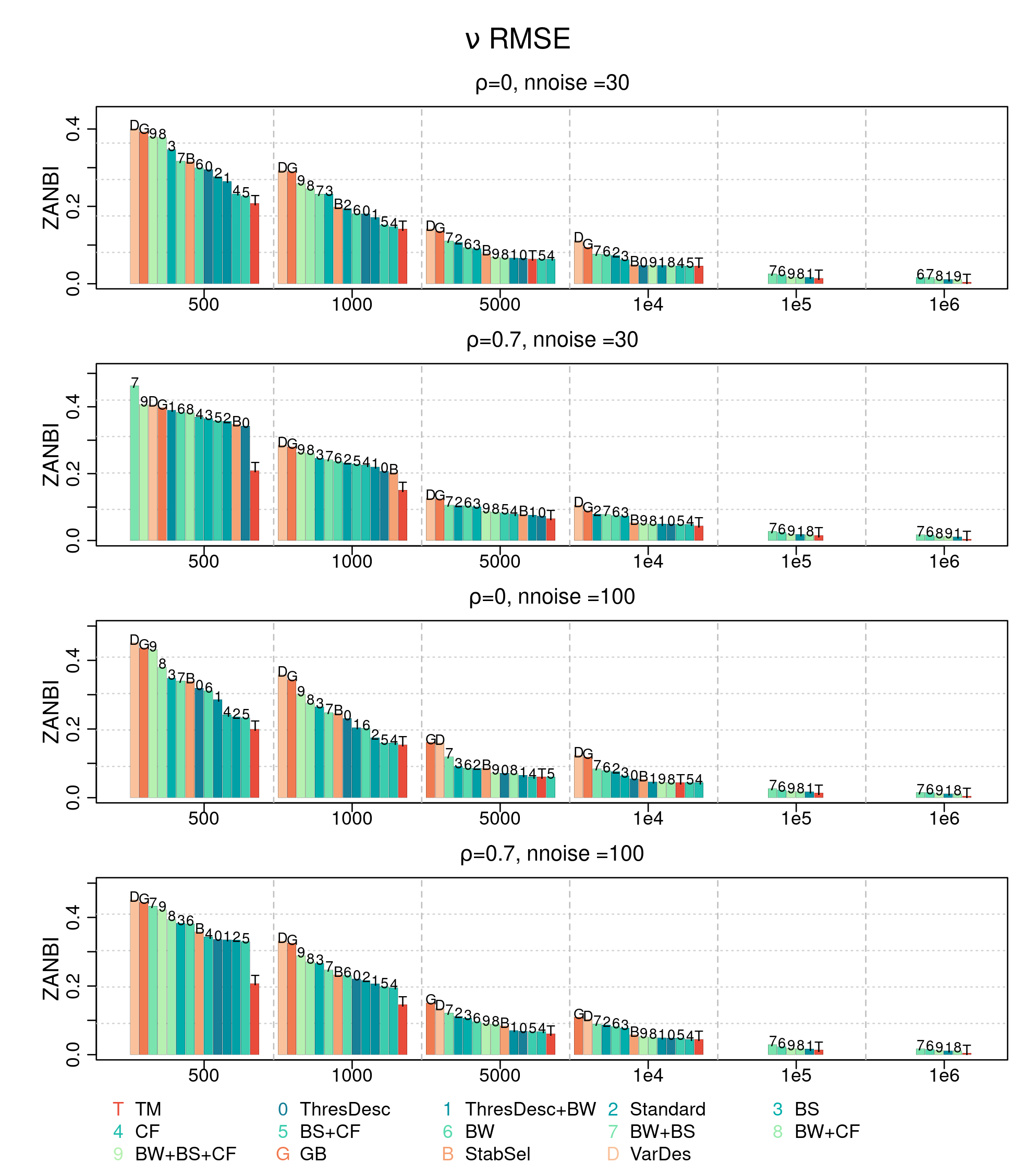}
\caption{\label{fig:nu} Simulation study. Average RMSE of the $\nu$-predictor of 100 replications vs. number of
observations for the different settings.}
\end{figure}

\begin{figure}[t!]
\centering
\includegraphics[width=0.82\textwidth]{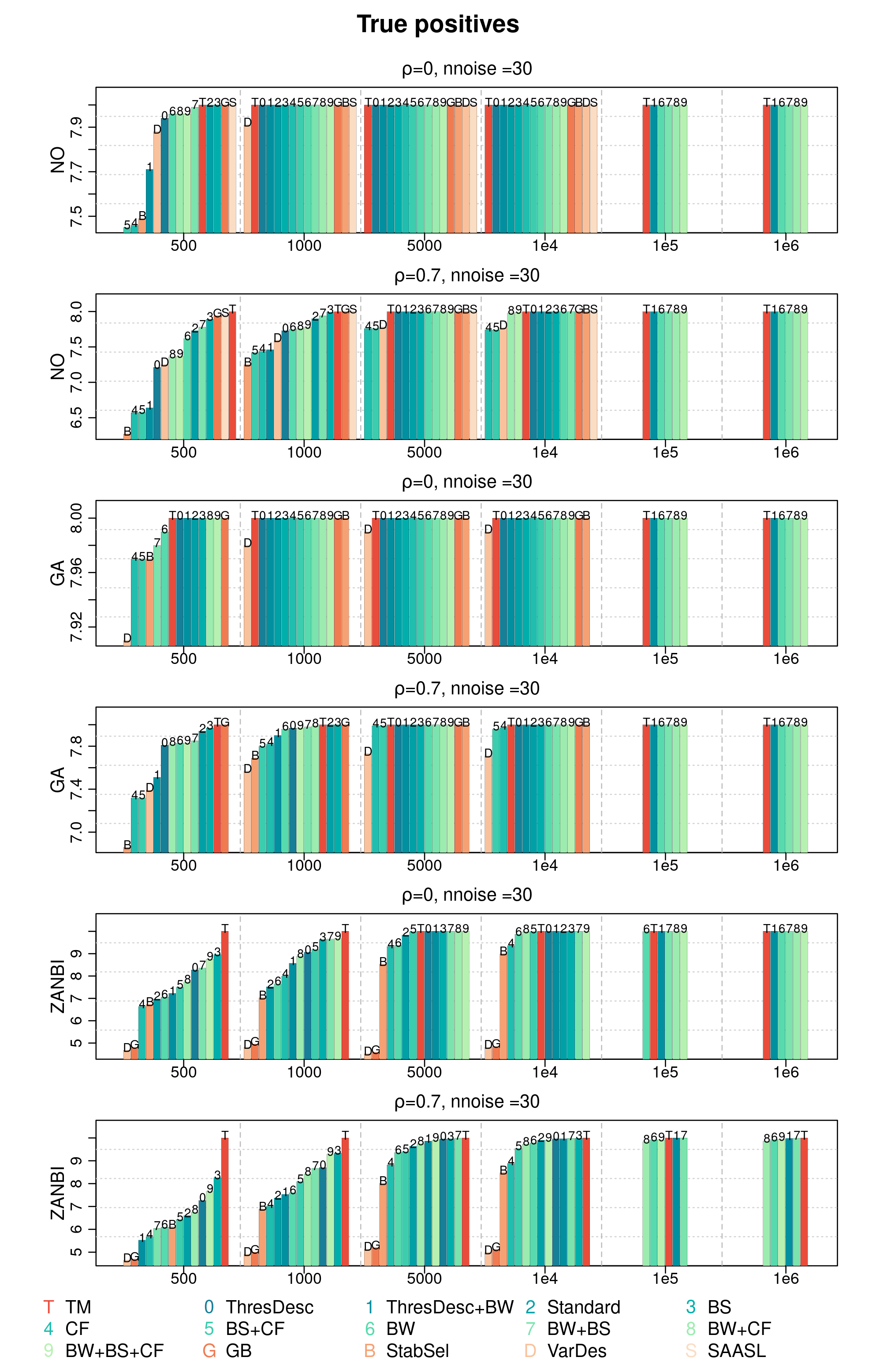}
\caption{\label{fig:tp1} Simulation study. Average true positive rate of 100 replications vs. number of
observations for the different settings.}
\end{figure}

\begin{figure}[t!]
\centering
\includegraphics[width=0.82\textwidth]{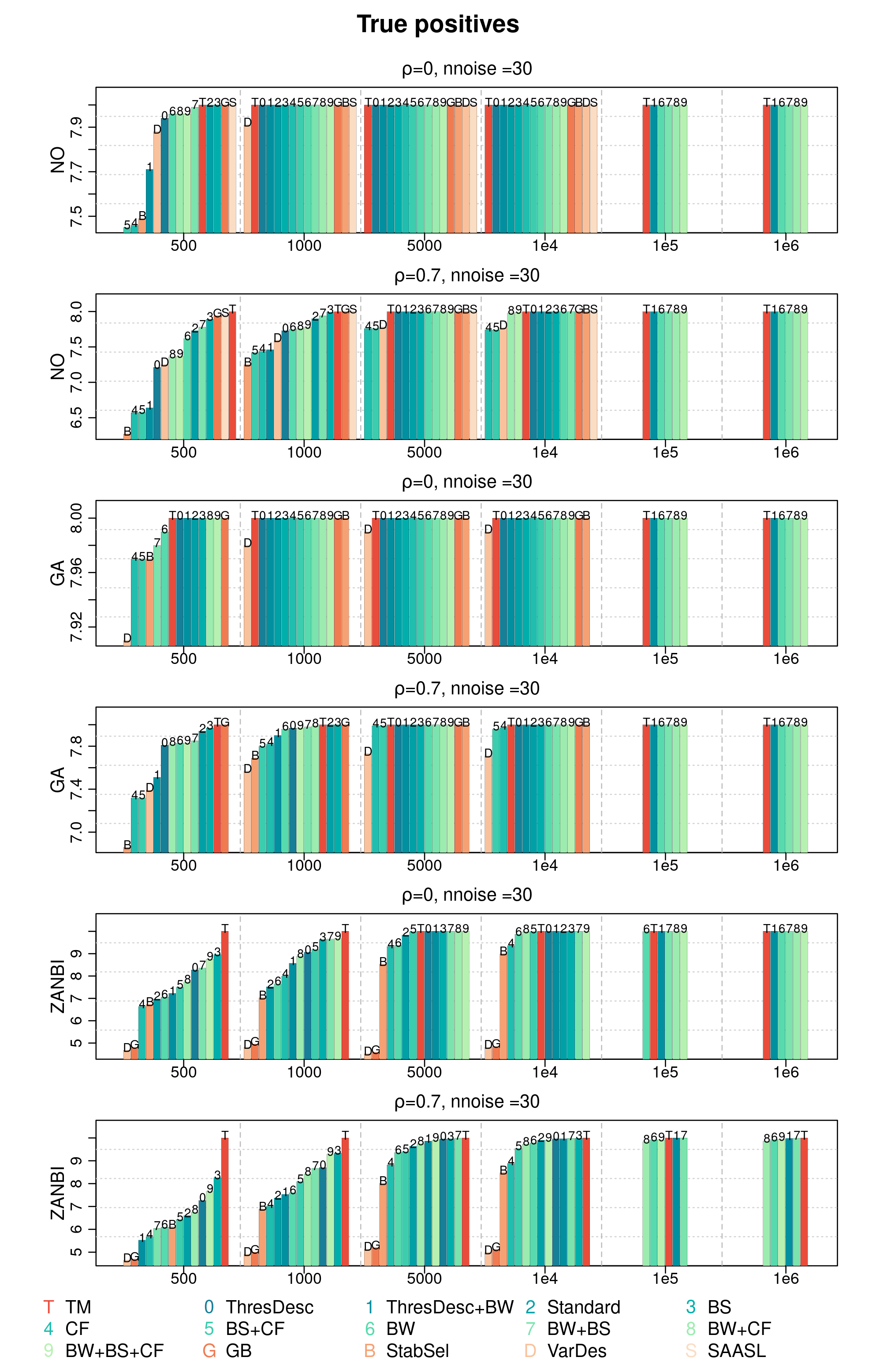}
\caption{\label{fig:tp2} Simulation study. Average true positive rate of 100 replications vs. number of
observations for the different settings.}
\end{figure}

\begin{figure}[t!]
\centering
\includegraphics[width=0.82\textwidth]{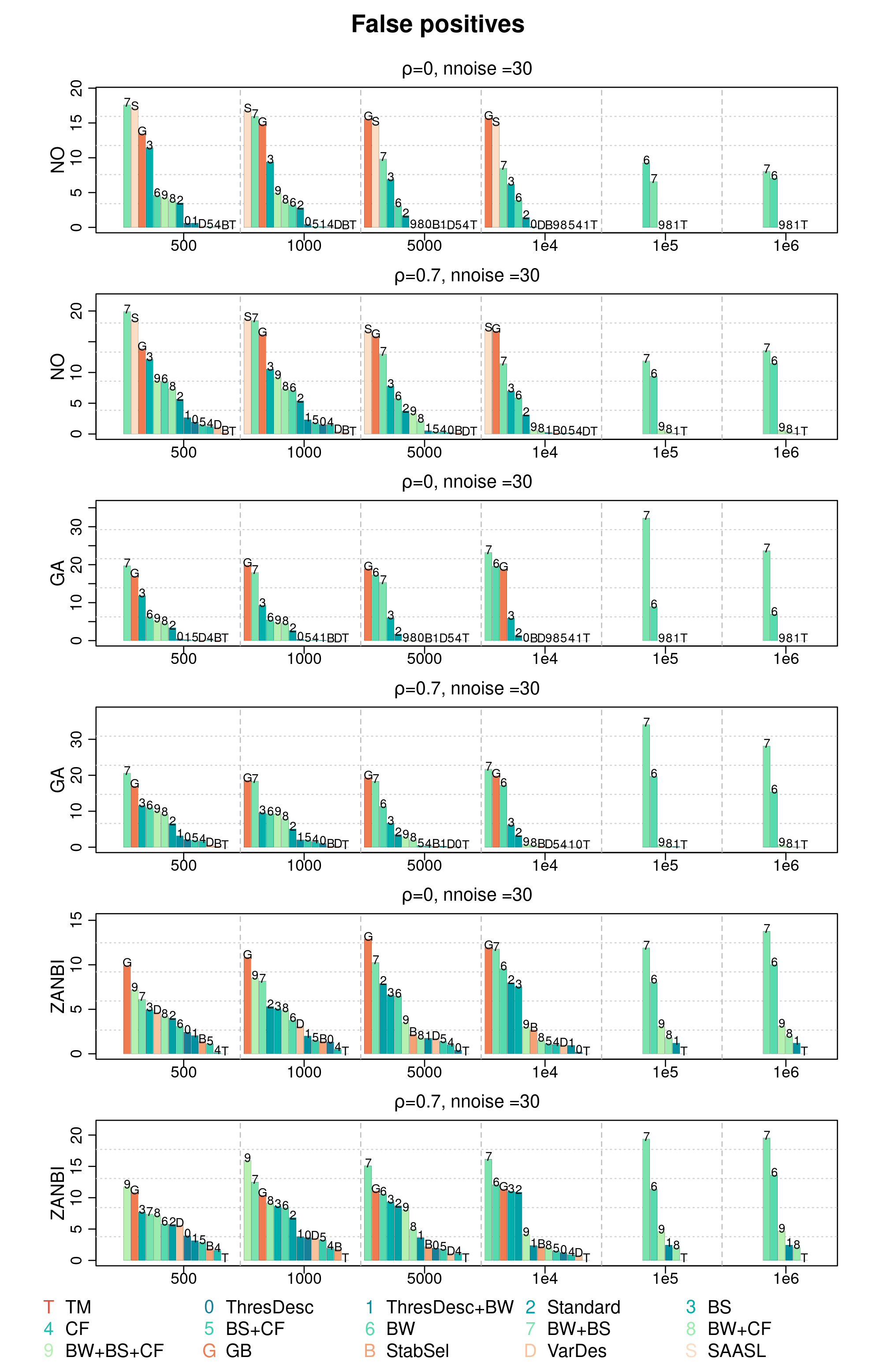}
\caption{\label{fig:fp1} Simulation study. Average false positive rates of $100$ replications vs.\ number
  of observations for the different settings.}
\end{figure}

\begin{figure}[t!]
\centering
\includegraphics[width=0.82\textwidth]{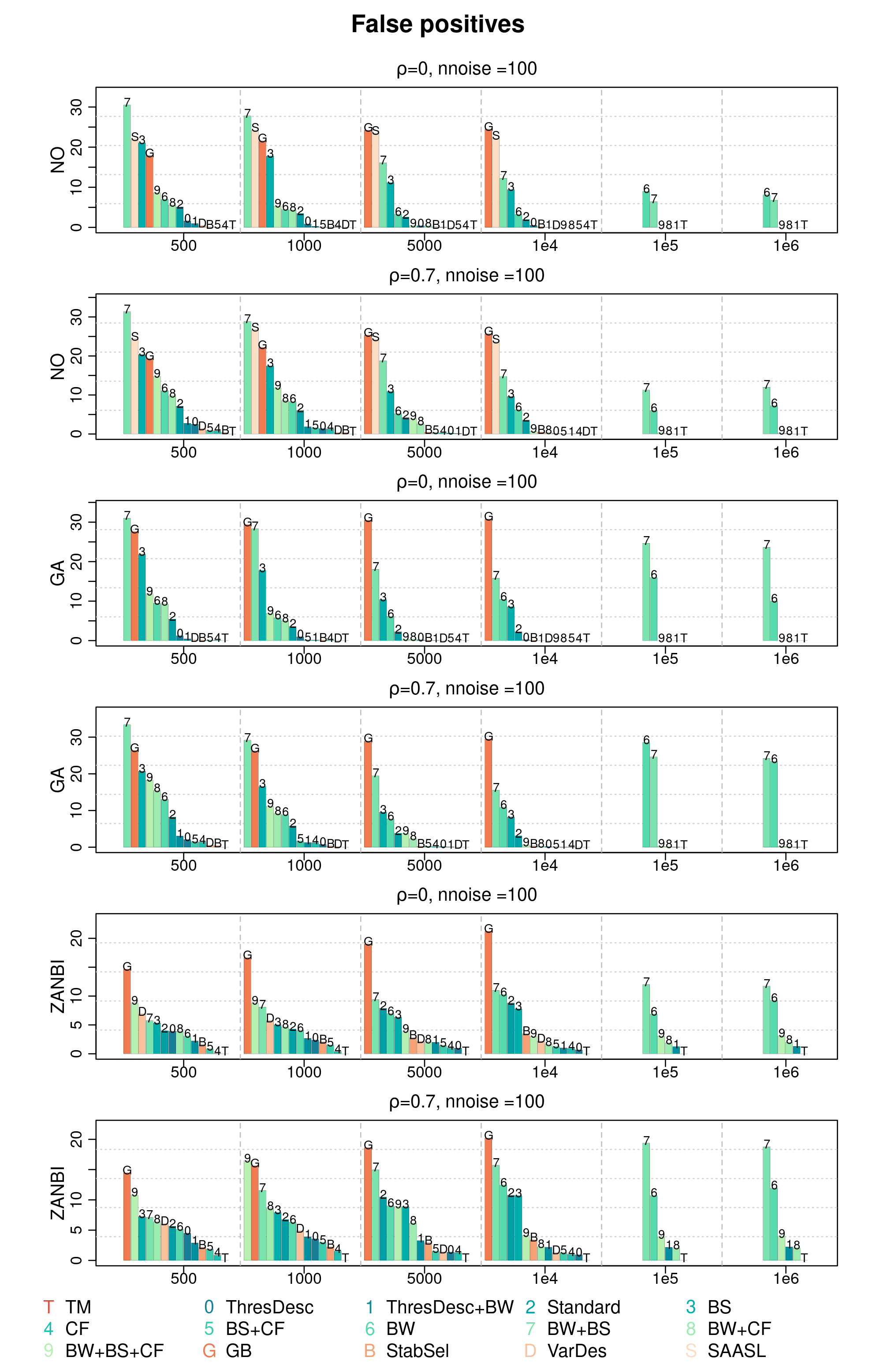}
\caption{\label{fig:fp2} Simulation study. Average false positive rates of $100$ replications vs.\ number
  of observations for the different settings.}
\end{figure}

\begin{figure}[t!]
\centering
\includegraphics[width=0.82\textwidth]{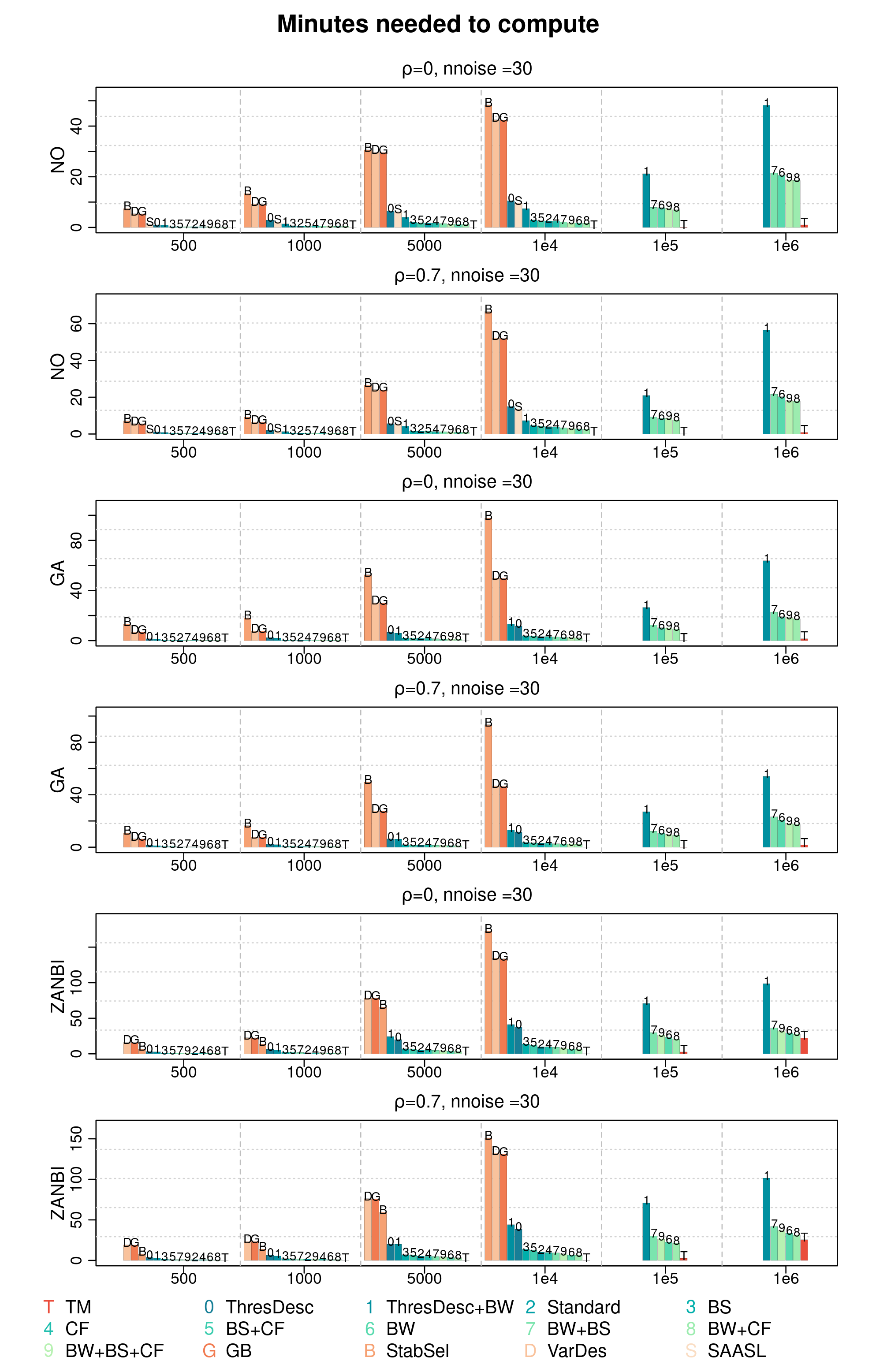}
\caption{\label{fig:elapsed1} Simulation study. Average time needed to compute a single model of $100$ replications vs.\ number
  of observations for the different settings.}
\end{figure}

\begin{figure}[t!]
\centering
\includegraphics[width=0.82\textwidth]{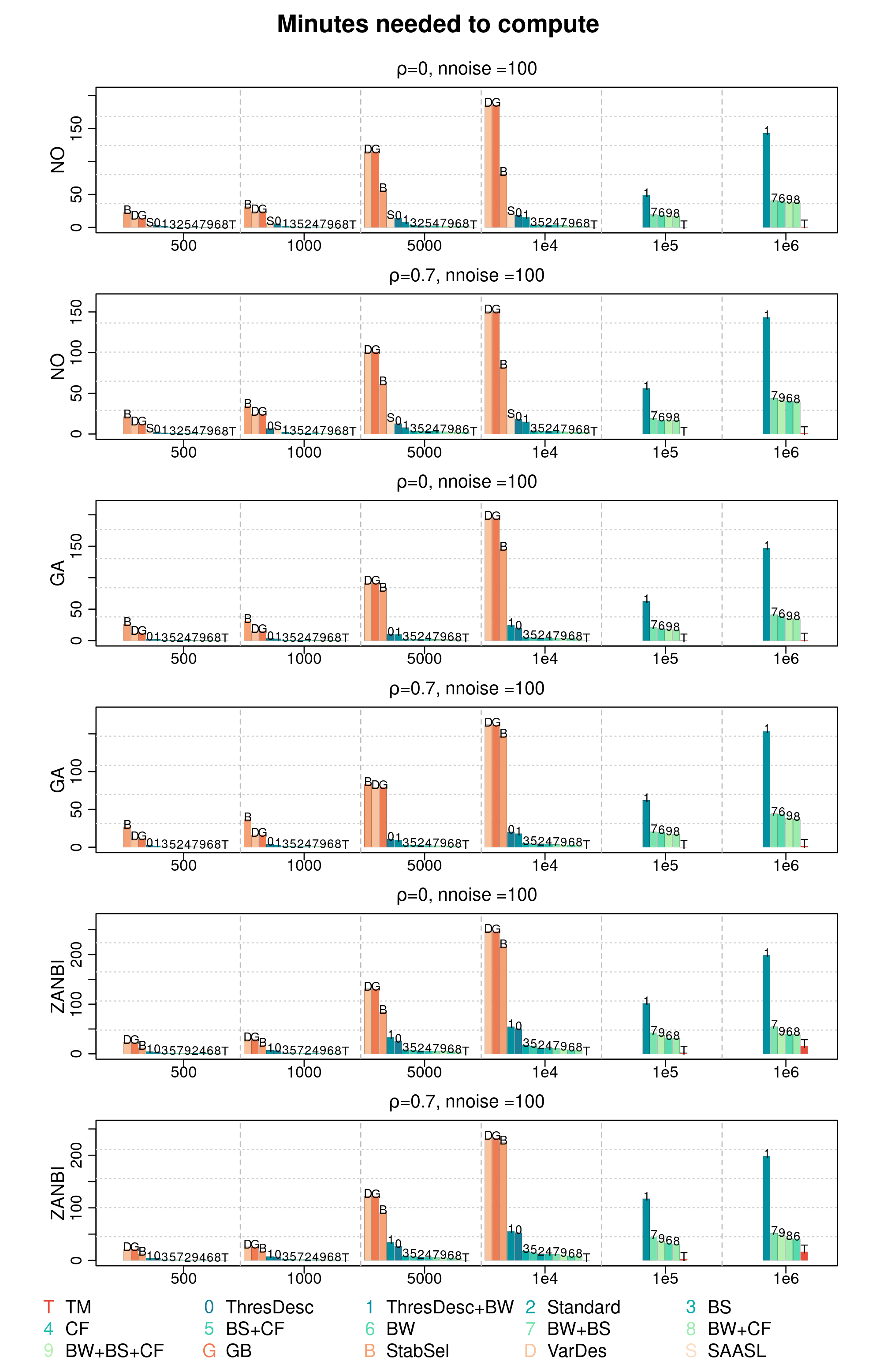}
\caption{\label{fig:elapsed2} Simulation study. Average time needed to compute a single model of $100$ replications vs.\ number
  of observations for the different settings.}
\end{figure}

\FloatBarrier

\end{document}